\documentclass[useAMS,usenatbib]{mn2e}

\usepackage{graphicx}

\title[Polarized broad emission lines]{The polarized signal from broad emmission lines in AGN}
\author[P. Lira et al.]{P. Lira$^{1}$, R. W. Goosmann$^{2}$, M. Kishimoto$^{3}$ and R. Cartier$^{4}$\\
$^{1}$Departamento Astronomia, Universidad de Chile, Casilla 36D, Santiago, Chile\\
$^{2}$Observatoire Astronomique de Strasbourg, Universit\'e de Strasbourg, CNRS, UMR 7550, 11 rue de l'Universit\'e, F-67000 Strasbourg, France\\
$^{3}$Department of Physics, Faculty of Science, Kyoto Sangyo University, Motoyama, Kamigamo, Kita-ku, Kyoto, 603-8555 Japan\\
$^{4}$Cerro Tololo Inter-American Observatory, Colina El Pino, Casilla 603, La Serena, Chile}
\begin{document}

\date{}

\pagerange{\pageref{firstpage}--\pageref{lastpage}} \pubyear{2002}

\maketitle

\label{firstpage}

\begin{abstract}

Using the {\it STOKES\/} Monte Carlo radiative transfer code we
revisit the predictions of the spectropolarimetric signal from a
disc-like Broad Emission Line Region (BLR) in Type I AGN due to
equatorial scattering. We reproduce the findings of previous works, but
only for a scatterer which is much more optically and geometrically
thick than previously proposed. We also find that when taking into
account the polarized emission from all regions of the scatterer, the
swing of the Polarizarion Angle (PA) is in the opposite direction to
that originally proposed. Furthermore, we find that the presence of
outflows in the scattering media can significantly change the observed
line profiles, with the PA of the scattering signal being enhanced in
the presence of radially outflowing winds. Finally, a
characteristically different PA profile, shaped like an `M', is seen
when the scatterer is cospatial with the BLR and radially outflowing.

\end{abstract}


\section{Introduction}

A characterization of the geometry and dynamics of the Broad Emission
Line Region (BLR) in Active Galactic Nuclei (AGN) has long been the
subject of intensive research. Some direct observations have now been
made by infrared interferometry on a single target (GRAVITY
collaboration et al. 2018) and, more generally, there is mounting
evidence that the BLR is likely a flattened system in Keplerian orbits
around the central black hole, with a possible contribution from a
wind component which seems to be more significant in high ionization
lines (Mej\'ia-Restrepo et al.~2018, Shen \& Ho 2014, Pancoast et
al.~2014, Runnoe et al.~2013, Proga \& Kurosawa 2010, Eracleous \&
Halpern 2003, 1994, Kollatschny 2003, Murray \& Chiang 1997, Chiang \&
Murray 1996, Marziani et al.~1996, Collin-Souffrin et al.~1988, Wills
\& Browne 1986).

Spectropolarimetry can be an extremely useful tool when addressing the
geometry and dynamics of the BLR. This technique adds two new
observables to those already available from the analysis of direct
light: the degree ($p$) and on-sky position angle (PA) of the
polarization of the line and continuum emission which gives
information about the scattering processes close to the central engine
{\em as a function of velocity}. Spectropolarimetry can also give an
indirect view of the observed system: that seen by the polarizing
material if the polarization is due to scattering. This was clearly
demonstrated over 35 years ago by the confirmation that type II AGN
(i.e., those where we do not have a direct view of the BLR) can show
broad components in their Balmer lines when viewed in polarized light
(e.g., Antonucci 1993 and references therein). In these sources the PA
is usually perpendicular to the axis of symmetry of the central
engine, implying that the light is polarized through scattering from
electrons or dust particles located above and/or below the central
region.

In a seminal work by Wood, Brown \& Fox (1993), changes in polarized
flux and PA were studied for the case of Thompson scattering from a
disc around line-emitting Be stars. Astonishingly, their analytic work
presents, to first approximation, many results with similar line
profiles to those presented here, although not always due to the same
geometric and dynamical considerations\footnote{As we will see,
  while in our case the line-emitting region, the BLR, corresponds to
  a disc-like structure undergoing Keplerian rotation, for Wood, Brown
  \& Fox (1993) the line emitting region, a Be star, corresponds to a
  static source. Hence, in their case the dynamical modulation of the
  line emission is {\em solely\/} due to the movement of the
  scatterer.}.

Later, Smith et al.~(2002, 2004, 2005) published a series of papers
looking at spectropolarimetric data for type I AGN (i.e., those where
we have a direct view of the BLR) and showed that the continuum
polarization is usually parallel to the axis of symmetry of the
systems, suggesting that the scattering material must be located in
the equatorial plane of the central source (as first pointed out by
Brown \& McLean 1977 and Shakhovskoi 1965).

The data also showed that the PA `swings' across the Balmer lines
which can be explained if the BLR has a disk-like geometry and the
scattering region is close enough to the BLR to spatially `resolve' it
(this is, that the red and blue Doppler-shifted wings of the emission
line are scattered at characteristically different PAs, as viewed by
the scatterer).

This understanding of how the inner AGN geometry shapes the
polarization across the broad emission lines opens up new
possibilities to constrain fundamental parameters of the system. A
recent study in this sense was published by Savi\v{c} et al.~(2018),
exploring to what extent the BLR polarization can detect Keplerian
motion around supermassive black holes and constrain their mass.

The Smith et al.~modelling, however, was based on semi-analytical
polarization code which considered only one scattering event per line
photon. Fast forward 15 years and computers are powerful enough to
determine the full Monte Carlo radiative transfer of the continuum and
line emission for the different physical, dynamical and geometric
conditions of the emitting and scattering regions. In this paper we
want to revisit the Smith et al.~results and expand their modelling
using a physically motivated parameter space of the BLR region and the
scattering media around it. In particular, one of the most interesting
outcomes, not explored by Smith et al., is found for a coincident
BLR/scatterer in the presence of an equatorial wind. Since nuclear
winds are thought to be a key component in many AGN, observationally
and theoretically, these results open the possibility to study such
winds using spectropolarimetry. In fact, we will use such results in
an upcoming publication to put real constraints on spectropolarimetric
observations of type I AGN and in this way extend our knowledge of the
physics of the BLR.

This paper is organized as follows: Section 2 presents the {\it
  STOKES\/} Monte Carlo code; Section 3 presents our rendition of the
Smith et al.~(2005) modelling; Sections 4, 5 and 6 present variations
and extensions to this paradigm; finally, Section 7 presents our
discussion and conclusions.

\begin{figure}
\includegraphics[scale=0.40,trim=50 0 0 100]{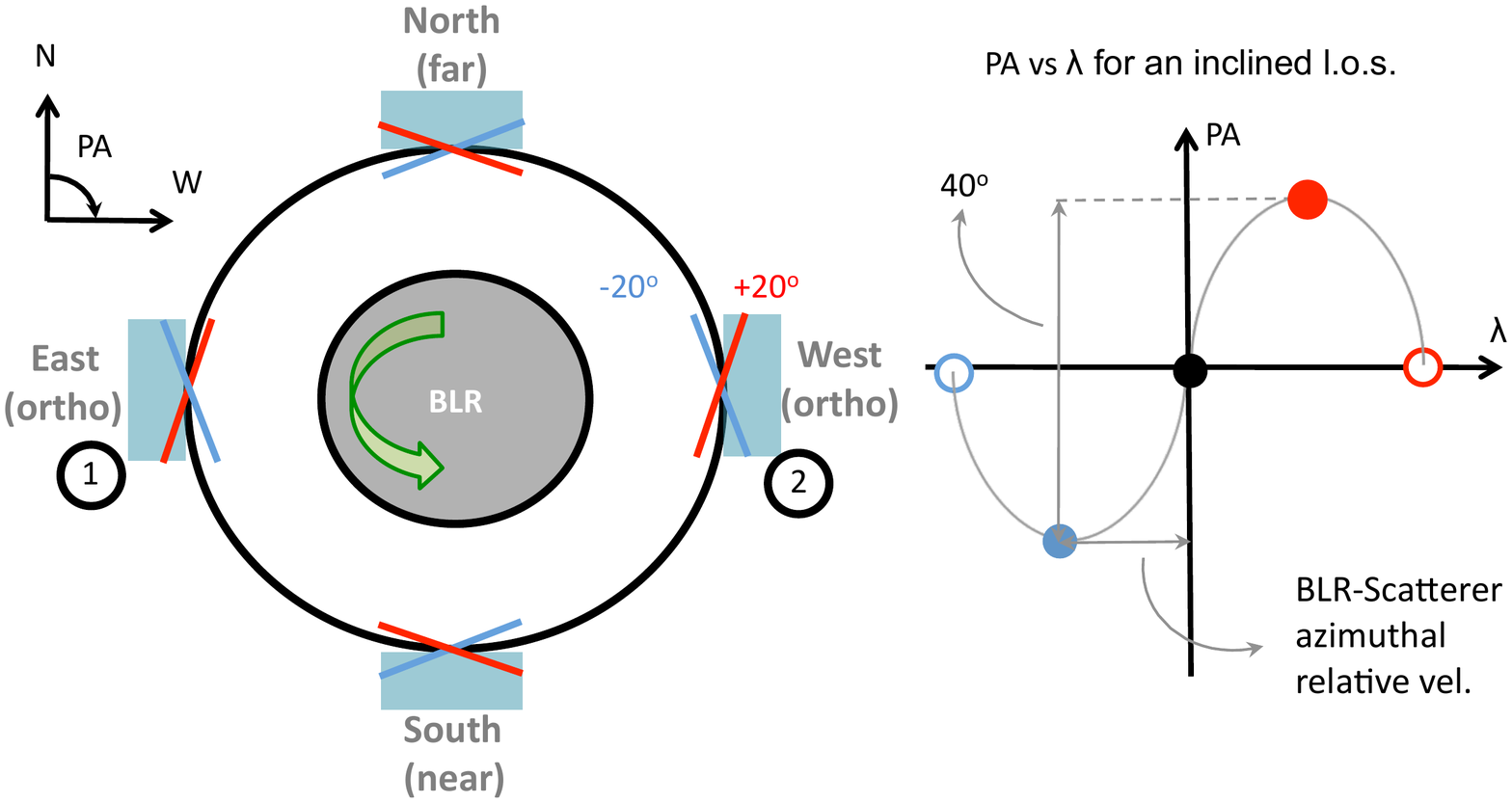}
\includegraphics[scale=0.30,trim=-50 0 100 200]{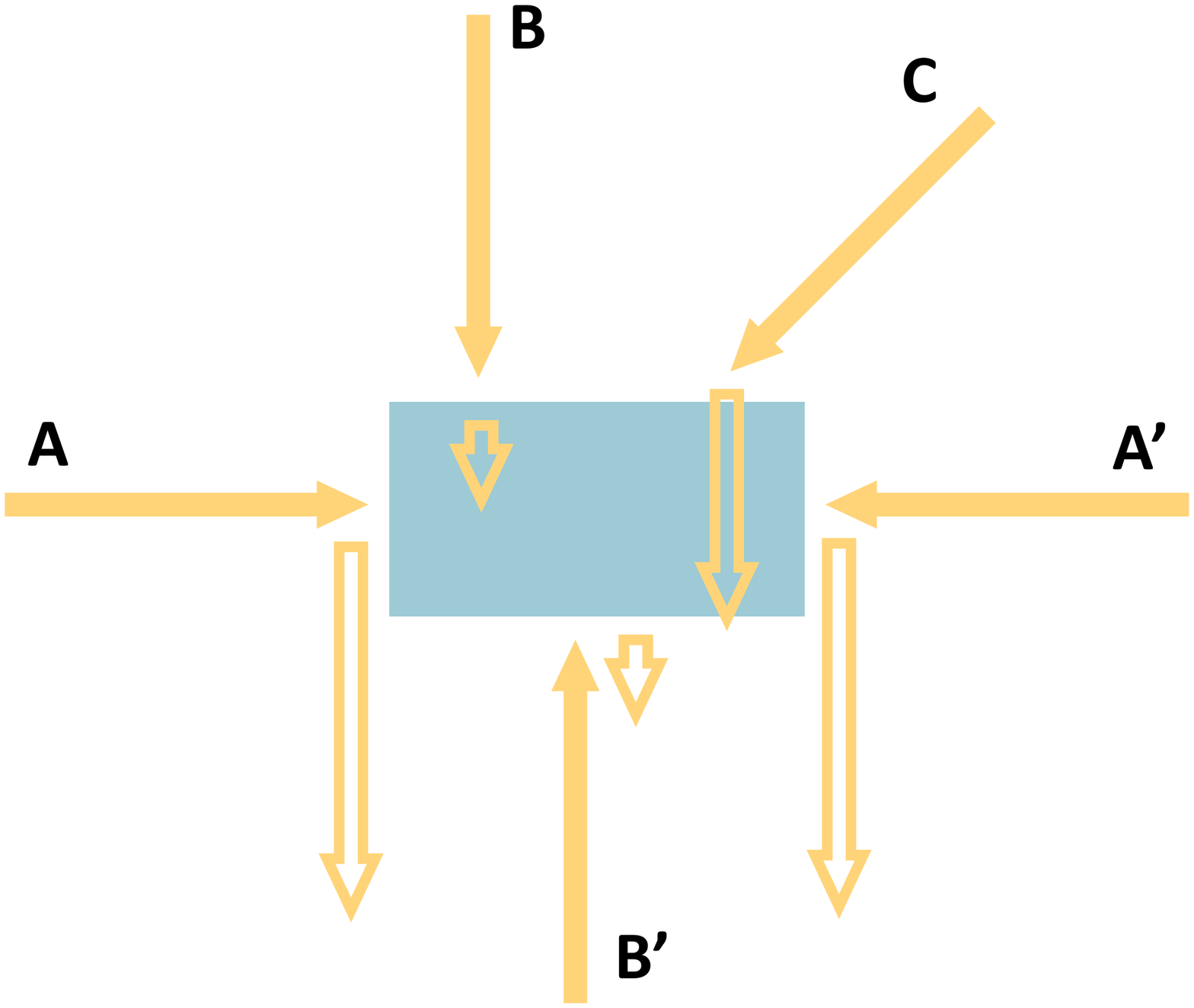}
\caption{{\bf Top:} Schematic representation of an anti-clockwise
  rotating BLR and the inner wall of the scattering region following
  S05. Only the BLR undergoes Keplerian rotation (green circular
  arrow), while the scatterer is at rest. The PA is measured West from
  North (i.e., clockwise) and the observer is located at a very large
  distance in the southern direction. The near, far and orthogonal
  locations in the scatterer, as seen by the observer, are shown. At
  each point of the scatterer inner wall, scattering elements see the
  velocity resolved BLR line emission with a difference of $\Delta$PA
  $= \pm 20^{o}$ (red and blue lines, with the black circle also
  showing the PA of light coming from the region straight ahead, where
  the continuum source -- the accretion disc -- is located). S05
  predicted that for inclined lines of sight, polarization from the
  orthogonal regions dominates the polarized signal (regions 1 \& 2),
  as the near and far sides are seen progressively more `in
  transmission', yielding a net PA $= \pm 20^{o}$ for large
  inclinations (see bottom figure), as represented in the spectrum
  presented on the right. Our {\it STOKES\/} realizations show that
  this is {\em not\/} the case. {\bf Bottom:} Representation of the
  polarization degree due to scattering events arriving at a
  scattering element (blue box) from different directions (A, B and
  C). The observer is coplanar with the scattering element and
  situated at the bottom of the figure, while the size of the (empty)
  arrows after the scattering event shows the level of polarization of
  the signal. For photons arriving from the sides of the scattering
  element (A), the polarization degree as measured by the observer
  will be maximum. For photons arriving from the top and bottom (B),
  the polarization degree will be minimum -- we call this
  configuration `in transmission'. For intermediate directions of the
  arriving photons (C), intermediate levels of polarization will be
  seen. Configuration A can explain the level of polarization expected
  from the orthogonal regions of the scatterer, while configuration B
  explains the level of polarization expected from the near and far
  sides.}
\label{schematicfigure}
\end{figure}

\section{The {\it STOKES\/} Monte Carlo code}

\begin{figure*}
\includegraphics[scale=0.85,trim=0 0 0 0]{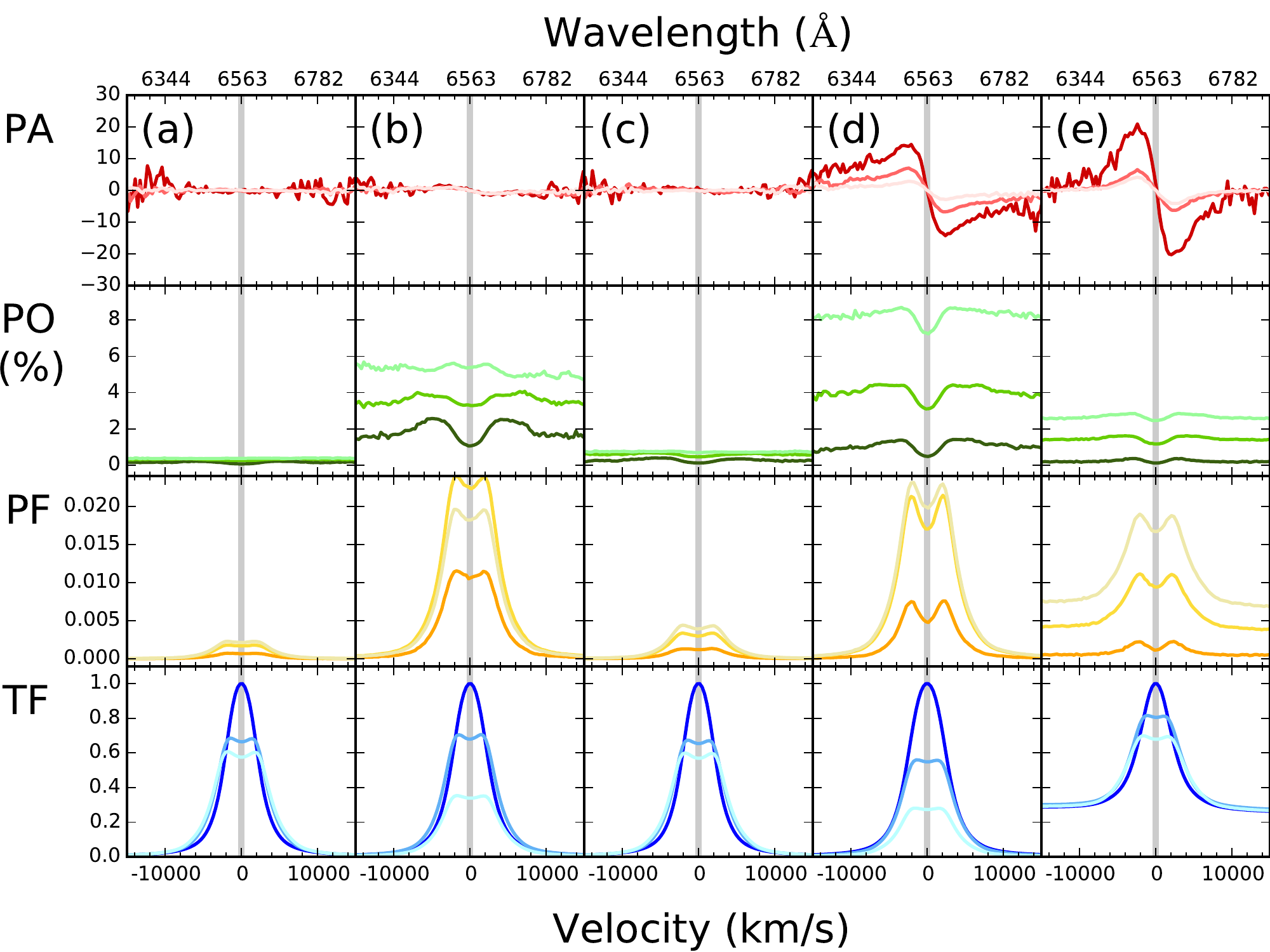}%
\caption{{\it STOKES\/} modelling following the Smith et al.~(2005)
  paradigm for a static equatorial outer scatterer with an electronic
  density of $1\times 10^6$ cm$^{-3}$ and a height of 0.001 pc (a),
  and an electronic density of $3\times 10^7$ cm$^{-3}$ and a height
  of 0.01 pc (b). Panels (c) and (d) show the results for the same set
  of electronic densities but for a scatterer height of 0.03 pc. Panel
  (e) shows modelling for the same BLR and scatterer parameters as (d)
  but including the contribution from the continuum unpolarized
  emission of a central source, with the line flux corresponding to
  40\%\ of total flux in the 5800--7200\AA\ range at the innermost
  radius. In each Panel, from top to bottom, we show the position
  angle (PA), percentage polarization (PO), polarized flux (PF) and
  total normalized flux (TF). The three models shown correspond to
  viewing angles of 24 (dark shades), 41 (medium shades) and 54 (light
  shades) degrees as measured from the axis of symmetry of the
  system. Smaller angles yield lower levels of polarization and larger
  PA changes. The thick grey continuous line corresponds to 0 km/s
  (6563\AA). Notice that subsequent figures do not have the same
  dynamical ranges in the y-axis.}
\label{modelS05}
\end{figure*}

We use version 1.2 of the Monte Carlo radiative transfer code {\it STOKES\/}
presented in Goosmann \& Gaskell (2007) and upgraded by Marin et
al.~(2012). This modelling suite coherently treats three-dimensional
radiative transfer and multiple reprocessing between emitting and
scattering regions and includes polarization. The system is surrounded
by a spherical web of virtual detectors. The detectors record the
wavelength, intensity and polarization state of each photon. The
latest version of {\it STOKES\/} also generates polarization images with the
photons being projected onto the observer's plane of the sky and then
stored in planar coordinates. The net intensity, polarization degree
$p$ and polarization position angle PA as a function of wavelength are
computed by summing up the {\it STOKES\/} vectors of all detected photons in a
given spectral and spatial bin. The spectra can be evaluated at each
viewing direction in the polar and azimuthal directions. Note that in
this work a PA value equal to zero denotes a polarization state with
the E-vector oscillating in a direction parallel to the projected axis
of symmetry of the system, while for PA = 90 degrees the E-vector is
perpendicular to the projected axis\footnote{This is the same
  convention adopted by Smith et al~(2005), but note that it is {\em
    not\/} the default output of the {\it STOKES\/} code, where an angle of
  zero degrees corresponds to perpendicular polarization.}. The PA
{\em increases clockwise}, i.e., West from North.

In the following Sections we will describe our modelling of the BLR
and the scattering media. The dynamical parameterization will be done
using cylindrical (v$_{\rho}$, v$_{\phi}$, v$_{z}$) coordinates. Note
also that our BLR is not opaque to photons, i.e., photons can freely
cross from one hemisphere to the other. This is in line with the idea
of the BLR as a clumpy structure with a small filling factor.

\section{The S05 paradigm}

The basic set up of parameters in the Smith et al.~paradigm can be
seen in Table 1 of Smith et al.~(2005, hereafter S05). We assume a
central black hole of $3.5 \times 10^7$ M$_{\sun}$ in mass (compared
with a $4.2 \times 10^7$ M$_{\sun}$ in S05), surrounded by a Keplerian
thin disk BLR (that we assume to have a height of 0.001 pc, as no
explicit information is given in S05) with inner and outer radii $3
\times 10^{-3}$ and $3 \times 10^{-2}$ pc. The line emission generated
by this BLR is centered at 6563\AA\ and has an intrinsic width of
50\AA\ or 2286 km/s. A BLR emissivity falling as $\sqrt R$ is
assumed. The dynamics of the BLR are modelled as nested rings of width
0.001 pc each, for which a Keplerian azimuthal velocity (v$_{\phi}$)
is obtained as v$_{\phi}^2 = GM/R$, with $G$ the gravitational
constant, $M$ the black hole mass and $R$ the radius of the ring.

The scattering region is also thin (again we assume 0.001 pc) and has
an annular geometry with inner and outer radii of 0.045 and 0.072 pc,
and an initial number density of $10^6$ electrons per cubic
centimeter, as assumed in S05 (but see below). When a non-stationary
scattering medium is assumed, its dynamics --- as with the BLR --- are
modelled as nested rings of width 0.001 pc each with the necessary
velocity vectors applied.

{\it STOKES\/} randomly generates isotropically emitted photons in the
central continuum source and the BLR, and then follows them as they
move freely or are scattered in electron filled regions. Once the
photons escape all scattering regions they are registered by the
virtual detectors that surround the full $4\pi$ space.

With this set of parameters S05 proposed that for an inclined line of
sight (to avoid total cancellation of the polarized signal due to
symmetry) the PA of the emission lines `swings' as a function of line
velocity with respect to the continuum PA level, with the blue/red
wing being above/below (or vice-versa) the continuum level and
crossing it at the line center (i.e., at zero velocity), where the
velocity vector of the near side of the BLR-disk has no radial
component and is located in the same direction as the continuum source
(see Figure \ref{schematicfigure}). The wings of the PA profile
correspond to those regions with the largest wavelength shifts due to
the maximum observable BLR velocity projected along the line of sight,
as seen by the scattering elements and located at the innermost of the
BLR disk with a PA angle close to zero (depicted as empty red and blue
circles in Figure \ref{schematicfigure}). Note that S05 did not
include the continuum emission as part of their modelling and only
considered the location of the central source at the origin of the
coordinate system to determine the continuum PA with respect to that
of the emission line.

\subsection{General remarks}

As already found by S05, Figure \ref{modelS05} shows that the degree
of polarization (PO spectrum) increases with inclination angle
($\phi$), as the system appears less and less symmetric to the
viewer. In fact, for a pole-on view of the system ($\phi = 0$), no
polarization is expected as at each wavelength the opposite
polarization signal is produced at the opposite side of the
disc. However, as soon as the observer has an inclined view of the
system, polarization at the near and far sides of the scattering disc
should start to diminish as $\cos \phi$ as we see these photons more
and more in transmission (which are not polarized); the signal becomes
dominated by the polarization from the orthogonal regions of the
disc. Our simulations, however, show that while the polarization is
indeed dominated by the orthogonal regions, the situation is not
symmetric for the near and far sides of the disc due to the thickness
of the scatterer, with the far side being more depolarized than the
near side because photons have a higher probability of suffering
multiple scattering events.

At the same time, the amplitude of the PA swing becomes smaller for
more inclined systems, as the change of the sky-projected position
angle decreases. This is represented by a PA swing of $\pm 20$ degrees
around zero degrees at the orthogonal regions of the disc in Figure
\ref{schematicfigure}. Another noticeable result, pointed out by S05,
is that the line seen in polarized light (the PF spectrum) is broader
than that seen in direct light. This is due to the location of the
equatorial scatterer which maximizes the relative projected velocity
between the BLR and the scattering region. In other words, the
scattering region `sees' the full velocity field and produces a
broader, double-horn line, as expected for Keplerian line emitting
discs (e.g., Eracleous et al.~2009). This, when combined with the
direct emission (the TF spectrum), results in a high percentage
polarization (the PO spectrum) of the line wings and a dip at the line
center.  Finally, as can be seen, the TF spectra have been normalized
to the peak flux of the spectrum observed at the smallest inclination
angle (24 degrees), which corresponds to the largest solid angle of
the BLR as seen by the observer, yielding a larger total flux in the
line.

Panel (e) in Figure \ref{modelS05} includes the unpolarized continuum
emission from the central source with line emission corresponding to
40\%\ of the total flux in the 5800-7200\AA\ range at the innermost
BLR radius and decreasing outwards as the square root of the
radius. It can be seen that the main effects of adding the continuum
component are the presence of the polarized continuum in the PF
spectra and a decrease in the level of polarization across the
continuum and line in the PO spectrum due to dilution, making it
appear flatter (for a quantitative comparison between models see Table
2). The predicted PO values are very close to those typically observed
in type I AGN (e.g., Smith et al.~2002, Schmid et al.~2001, Smith et
al.~1997, Goodrich 1989). In addition, the line PA swing becomes more
pronounced and the very extended PA wings seen at high velocities are
no longer present. The explanation for this requires further
consideration.

The changes in the PA behaviour are not due to the geometric
configuration of the system, but to the vectorial nature of the
polarized signal\footnote{As before, notice that to derive the correct
  result the vectorial addition of the {\it STOKES\/} parameters is required,
  not that of the electromagnetic fields.}. The continuum emission is
always polarized at a PA equal to the tangential direction of the
circles depicted in Figure \ref{schematicfigure}.  At each wavelength
we need to add the coninuum and line emission, as vectors. But, as
already discussed, the PA of the line is constantly rotating as a
function of velocity and therefore the result of the vectorial
addition will vary with wavelength. The end result is that the PA
peaks, which {\em geometrically\/} correspond to the largest opening
angles seen by the scattering element (and therefore to intermediate
velocities of $\sim \pm 2000$ km/s -- see Figure
\ref{schematicfigure}), are increased by the addition of the continuum
vector, while the high velocity wings beyond $\sim \pm 10000$ km/s are
dominated by the zero angle continuum emission.

Since a model including a central emitting source is more realistic
than the original S05 scenario, all subsequent models will contain
both continuum and line emission. Parameters for the original S05
model, our realization, and subsequent models presented in the
following Sections, are given in Table 1.

\subsection{The optical depth and covering factor of the scatterer}

\begin{figure}
\includegraphics[scale=0.45,trim=0 0 0 0]{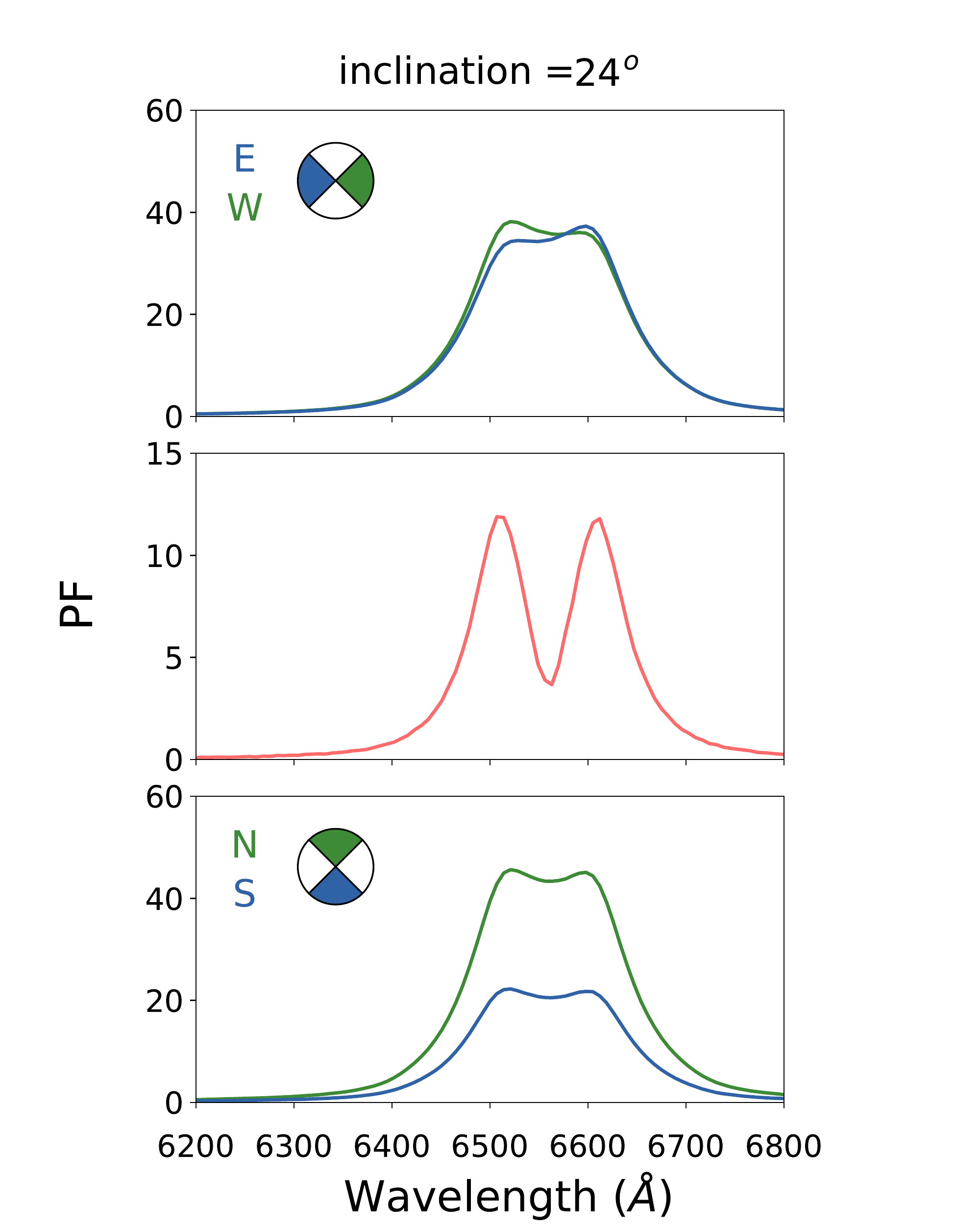}%
\caption{Polarized Flux (PF) as a function of wavelength for the four
  quadrants of the scattering region for model 2(d). These are the
  East, West (top panel), North and South (bottom panel) quadrants, as
  represented by the pie charts. The total spectrum is also presented
  (middle panel).}
\label{quad_PF_spectra}
\end{figure}

\begin{figure}
\includegraphics[scale=0.45,trim=0 0 0 0]{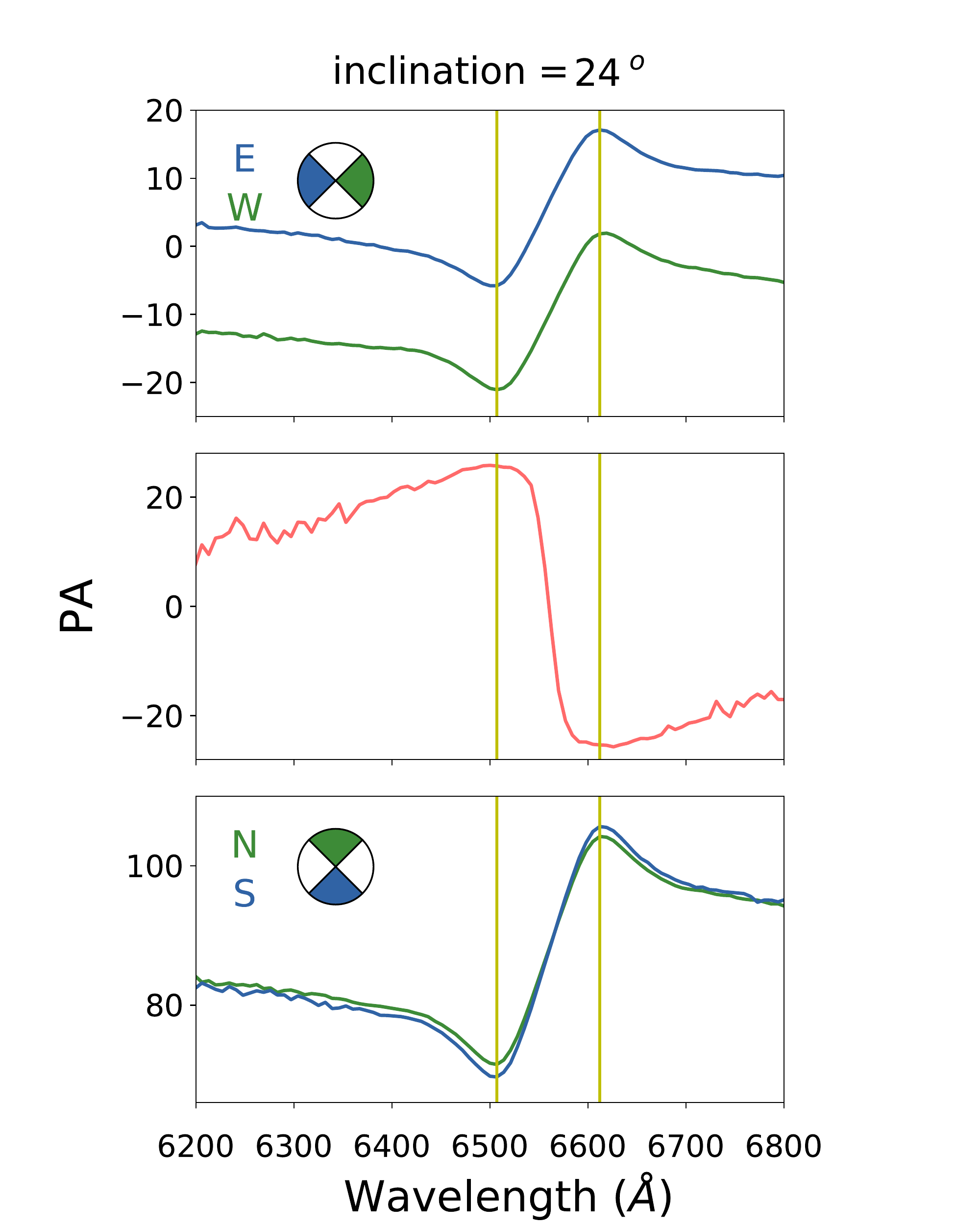}%
\caption{Position angle (PA) as a function of wavelength for the four
  quadrants of the scattering region for model 2(d). These are the
  East, West (top panel), North and South (bottom panel) quadrants, as
  represented by the pie charts. The total spectrum is also presented
  (middle panel).}
\label{quad_PA_spectra}
\end{figure}

\begin{figure*}
\includegraphics[scale=0.35,trim=0 0 0 100]{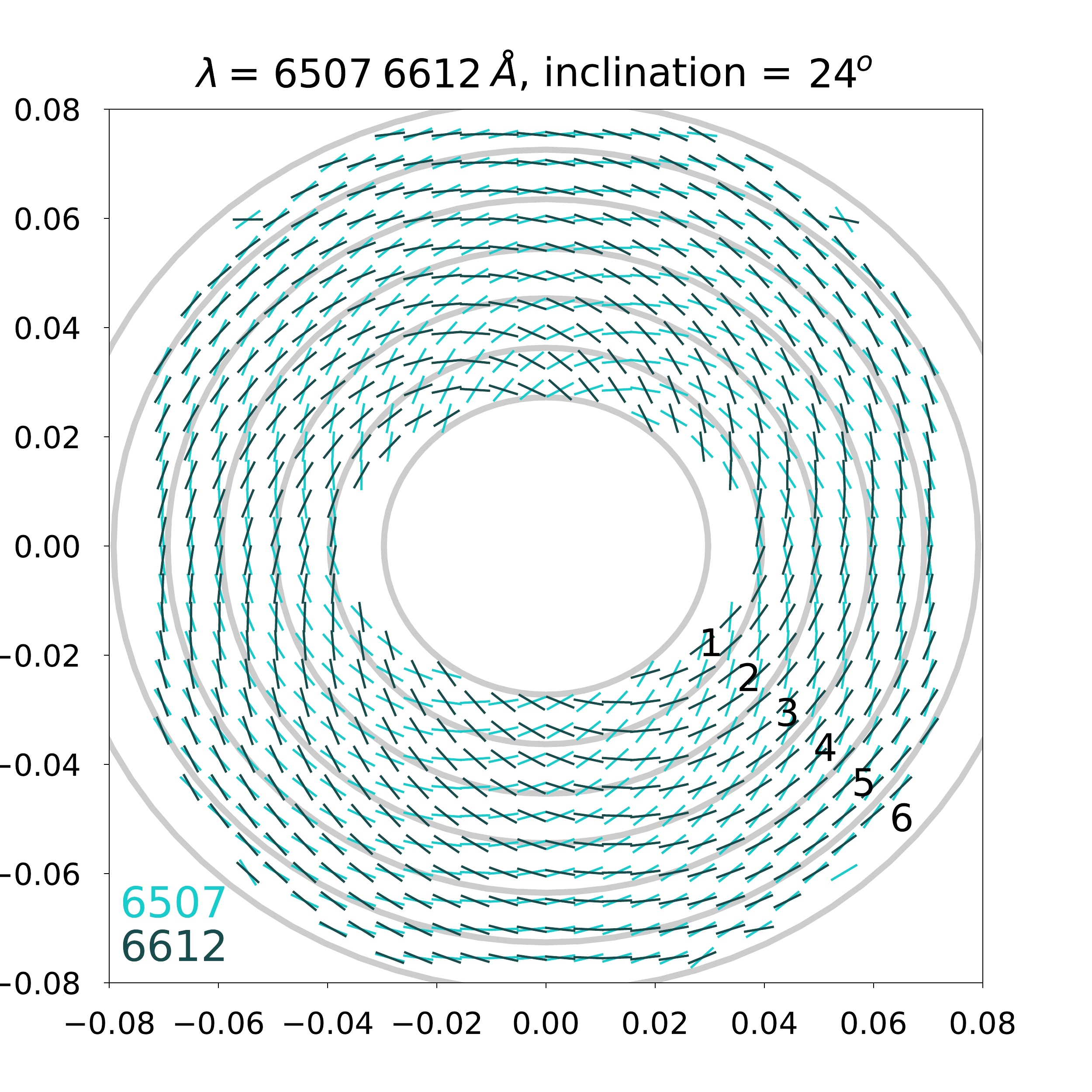}%
\includegraphics[scale=0.35,trim=0 0 0 100]{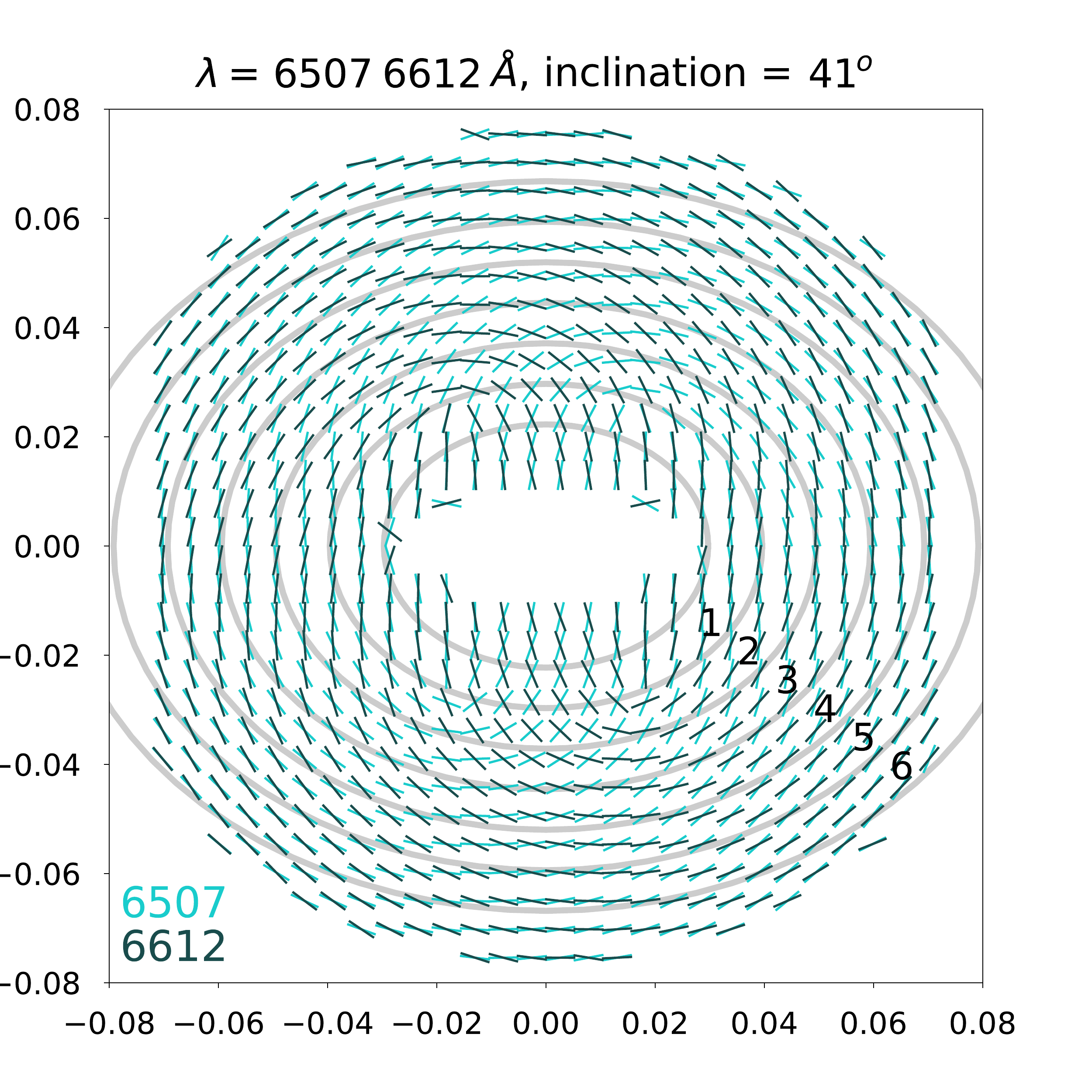}
\caption{PA spatially resolved maps of model 2(d) for 6507 \AA\ and
  6612 \AA\ at a viewing angle of 24 degrees (left) and 41 degrees
  (right). The markers show the PA orientation and their sizes are
  arbitrary. The scatterer has been divided into concentric
  projected-rings. The axes in all plots are expressed in pc.}
\label{pa_maps}
\end{figure*}

\begin{figure}
\includegraphics[scale=0.55,trim=0 0 0 0]{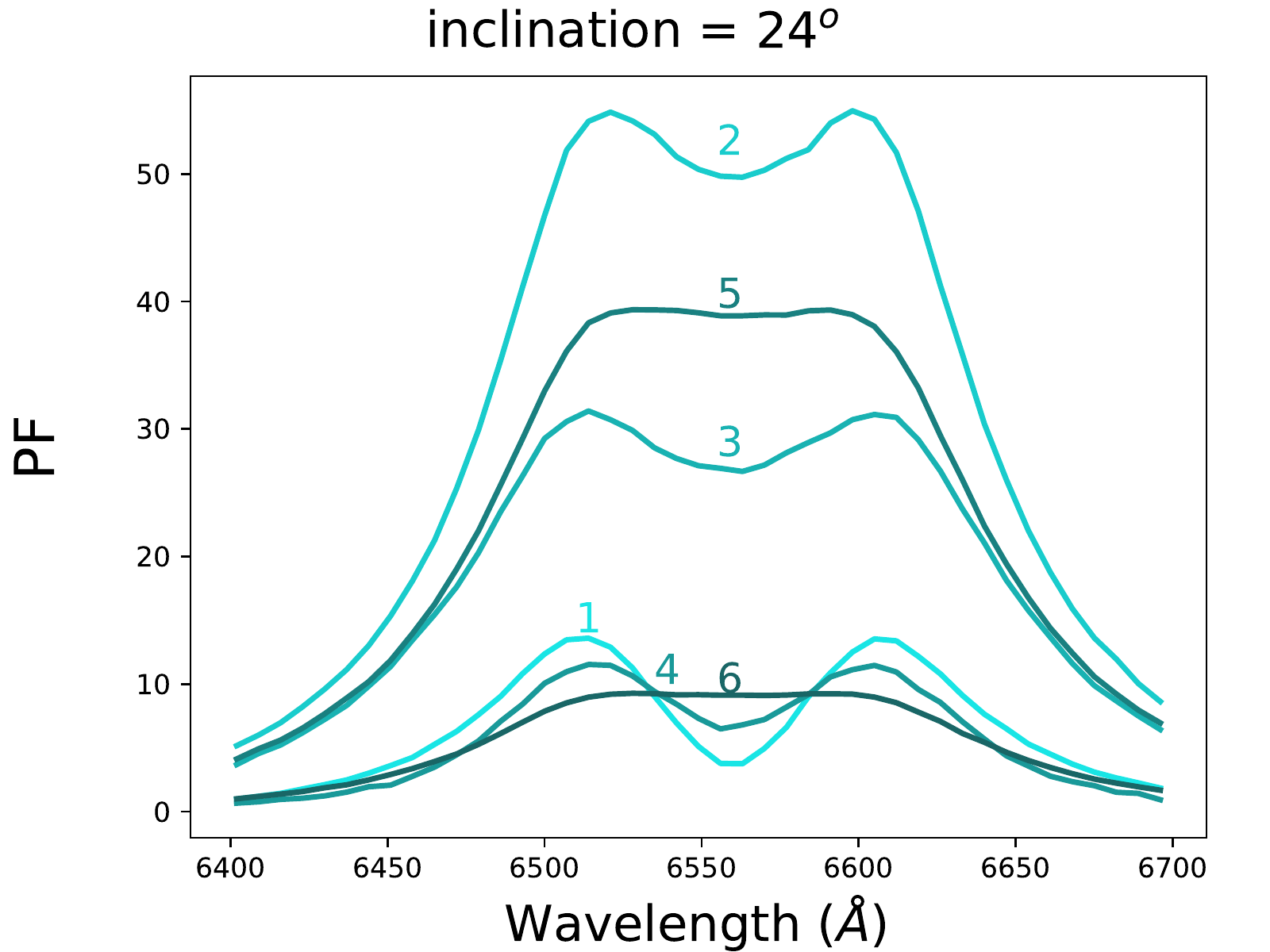}
\caption{PF spectra obtained from the spatially resolved maps of model
  2(d) for a viewing angle of 24 degress. The spectra correspond to
  the concentric rings represented in Figure 5.}
\label{ring_spec}
\end{figure}
 
One important element of the S05 paradigm is that the net
polarization, and therefore the resulting observed PA, will be
dominated by the orthogonal regions of the scatterer, as shown in
Figure \ref{schematicfigure}. Those regions found at the near and far
side of the scatterer, which are seen progressively more `in
transmission' for more inclined views, are mostly canceled out.

We started by computing the polarimetric signal for the original S05
parameter set up. However, we failed to reproduce their results and
instead found a very low degree of polarization ($< 0.3\%$) with no PA
change across the line (see Figure \ref{modelS05}, Panel (a)). It soon
became clear that a higher density and much larger covering factors
were needed to reproduce the line scattering signal. Trial and error
showed that electron densities of $\la 10^7$ cm$^{-3}$ had sufficient
optical depth to return higher levels of polarization, but still gave
a very low amplitude swing across the emission line (Figure
\ref{modelS05}, Panel (b)).  A dramatic increase in the solid angle of
the scatterer as seen by the BLR (from a total height of 0.001, to
0.01 and finally 0.03; see Table 1) is necessary to increase the
swing to amplitudes comparable to those observed in Seyfert 1 galaxies
(Fig \ref{modelS05}, Panel (d)). When comparing with S05 it can be
seen that the Monte Carlo code predicts a less sinusoidal-like PA
swing, with a slower rise/fall of the high velocity winds.

The large optical depths and covering factors imply an asymmetry
between the far and near side of the system.  Figure
\ref{quad_PF_spectra} shows the resulting PF spectra after averaging
over four different regions of the scatterer: the North, South, East
and West quadrants. While back-scattering off the far side (N) of the
scattering region gives a high level of polarized flux from photons
that can reach the observer, photons that undergo forward-scattering
while traveling towards the observer and across the scatterer in the
near side (S) yield about half the level of polarized flux. A small
difference is also seen in the strength of the blue and red horns
between the East and West quadrants. This is a significant departure
from the results found by S05.

\subsection{The sense of the PA swing}

Our simulation show that the swing in the PA spectrum is in the
opposite sense to that predicted by S05 and shown in Figure
\ref{schematicfigure}, with the blue side of the line seen at positive
PA values and the red side at negative PA values.  We explain this
inversion in the rest of this section.

Figure \ref{quad_PA_spectra} shows the PA spectra obtained from
averaging the {\it STOKES\/} vectors over the N, S, E and W quadrants. All
show the PA swing expected from Figure \ref{schematicfigure}. The E
and W quadrants average PA, however, is not located at 0 degrees. This
is because of the different levels of polarized flux coming from the
near and far side of the scatterer, with the emission from the far
side being much more dominant, as discussed in the previous
section. This biases the resulting average PA of the E and W quadrants
towards northern values. This effect goes away in simulations with
smaller optical depths.

The spectrum showing the PA obtained when averaging the {\it STOKES\/}
parameters over the full scattering region, however, reverts to what
was presented in Figure \ref{modelS05}. The nearly identical PA
signature coming from the N and S quadrants, which are affected by
very different optical depths, suggests that optical thickness is not
responsible for the reverse in the PA swing. So the sense of the swing
of the resulting PA should be due to geometric cancellation between
quadrants.

\begin{table*}
\caption{Geometric and dynamical parameters of the BLR and scattering regions used for the {\it STOKES\/} modelling.}
\begin{tabular}{lrrrrcrrrrrrrrr} \hline
Model & \multicolumn{4}{c}{Broad Line Region} && \multicolumn{9}{c}{Scatterer} \\ \cline{2-5} \cline{7-15}
& R$_{in}$ & R$_{out}$ & Height & v$_{\phi} \dag$ && R$_{in}$ & R$_{out}$ & Height & v$_{\rho}$ & v$_{\phi} \dag$ & v$_{z}$ & $\tau_{z}$ & $\tau_{r}$ & n$_e$ \\ 
&  pc     &  pc      &  pc    &  km/s          &&  pc     &  pc      &  pc    &  km/s  &  km/s   &  km/s      &           &            &  cm$^{-3}$ \\

\cline{1-1} \cline{2-5} \cline{7-15}
%                                                                               v_{rho} v_{phi} v_{z}       
&&&&&&&&&&&&&&\\                                                                                         
S05    & 0.003 & 0.035 & 0.001 & 6736               && 0.045 & 0.073 & 0.001     & 0   & 0     & 0       & -- & -- & 1e6 \\ 
&&&&&&&&&&&&&&\\                                                                                         
Fig2(a)& 0.003 & 0.035 & 0.001 & 6736               && 0.045 & 0.073 & 0.01      & 0   & 0     & 0       & 0.02 & 0.054 & 1e6 \\
Fig2(b)& 0.003 & 0.035 & 0.001 & 6736               && 0.045 & 0.073 & 0.01      & 0   & 0     & 0       & 0.6  & 1.62 & 3e7 \\
Fig2(c)& 0.003 & 0.035 & 0.001 & 6736               && 0.045 & 0.073 & 0.03      & 0   & 0     & 0       & 0.06 & 0.054 & 1e6 \\
Fig2(d)& 0.003 & 0.035 & 0.001 & 6736               && 0.045 & 0.073 & 0.03      & 0   & 0     & 0       & 1.8  & 1.62 & 3e7 \\
Fig2(e)$\diamondsuit$& 0.003 & 0.035 & 0.001 & 6736 && 0.045 & 0.073 & 0.03      & 0   & 0     & 0       & 1.8  & 1.62 & 3e7 \\
&&&&&&&&&&&&&&\\                                                                                       
Fig7(a)$\diamondsuit$& 0.003 & 0.035 & 0.001 & 6736 && 0.043 & 0.073 & 0.03      & 0   & 2030  & 0       & 1.8  & 1.8 & 3e7 \\
Fig7(b)$\diamondsuit$& 0.003 & 0.035 & 0.001 & 6736 && 0.043 & 0.073 & 0.03      & 0   & 2030  &$\pm$3000& 1.8  & 1.8 & 3e7 \\
Fig7(c)$\diamondsuit$& 0.003 & 0.035 & 0.001 & 6736 && 0.043 & 0.073 & 0.03      & 3000& 2030  & 0       & 1.8  & 1.8 & 3e7 \\
Fig7(d)$\diamondsuit$& 0.003 & 0.035 & 0.001 & 6736 && 0.043 & 0.073 & 0.03      & 2121& 2030  &$\pm$2121& 1.8  & 1.8 & 3e7 \\
&&&&&&&&&&&&&&\\                                                                                         
Fig8(a)$\diamondsuit$& 0.003 & 0.035 & 0.001 & 6736 && 0.003 & 0.035 & 0.001     & 0   & 6736  & 0       & 0.06 & 1.92 & 3e7 \\
Fig8(b)$\diamondsuit$& 0.003 & 0.035 & 0.001 & 6736 && 0.003 & 0.035 & 0.001     & 0   & 6736  &$\pm$3000& 0.06 & 1.92 & 3e7 \\
Fig8(c)$\diamondsuit$& 0.003 & 0.035 & 0.001 & 6736 && 0.003 & 0.035 & 0.001     & 3000& 6736  & 0       & 0.06 & 1.92 & 3e7 \\
Fig8(d)$\diamondsuit$& 0.003 & 0.035 & 0.001 & 6736 && 0.003 & 0.035 & 0.001     & 2121& 6736  &$\pm$2121& 0.06 & 1.92 & 3e7 \\
&&&&&&&&&&&&&&\\                                                                                         
\hline
\multicolumn{15}{l}{$\diamondsuit$: Continuum emission included.}\\
\multicolumn{15}{l}{$\dag$: v$_{\phi}$ at R $=$ R$_{in}$.}
\end{tabular}
\end{table*}

\begin{table*}
\caption{Continuum, minimum and maximum values observed in the PA, PO and PF spectra, and their differences. $\Delta v$ gives the difference between the velocities at which the minimum and maximum values are observed in km/s (i.e.,  $\Delta v = v_{\rm max} - v_{\rm min}$).}
\begin{tabular}{rrrrrrrrrrrrrrrrrr} \hline
& \multicolumn{5}{c}{PA} && \multicolumn{5}{c}{PO(\%)} && \multicolumn{5}{c}{PF} \\ \cline{2-18}
Model & cont & min & max & $\Delta$PA & $\Delta v$ && cont & min & max & $\Delta$PO & $\Delta v$ && cont & min & max & $\Delta$PF & $\Delta v$ \\ \cline{2-6} \cline{8-12} \cline{14-18}
&&&&&&&&&&& \\

Fig2(d) $\phi= 24^{\circ}$ &  5 & -14 & 14 & 29 & -4480 && 0.9 & 0.5 & 1.4 & 1.0 & 5120 && 0.000 & 0.001 & 0.008 & 0.007 & 9599 \\
$\phi= 41^{\circ}$ &  1 & -7 &  7 & 14 & -4800 && 3.8 & 3.1 & 4.4 & 1.3 & -6080 && 0.000 & 0.004 & 0.038 & 0.034 & 9279 \\
$\phi= 54^{\circ}$ &  0 & -3 &  3 &  6 & -4800 && 8.3 & 7.2 & 8.7 & 1.5 & -3520 && 0.001 & 0.009 & 0.082 & 0.073 & 5440 \\
Fig2(e) $\phi= 24^{\circ}$ &  1 & -20 & 21 & 41 & -4480 && 0.2 & 0.1 & 0.4 & 0.3 & -2560 && 0.001 & 0.001 & 0.002 & 0.002 & -3840 \\
$\phi= 41^{\circ}$ &  0 & -6 &  7 & 13 & -4800 && 1.4 & 1.2 & 1.6 & 0.5 & 3840 && 0.005 & 0.006 & 0.014 & 0.008 & 5120 \\
$\phi= 54^{\circ}$ &  0 & -4 &  4 &  8 & -5120 && 2.6 & 2.5 & 2.9 & 0.4 & 2880 && 0.011 & 0.013 & 0.027 & 0.014 & 5120 \\
Fig7(a) $\phi= 24^{\circ}$ &  0 & -30 & 30 & 60 & -1600 && 0.3 & 0.1 & 0.5 & 0.5 & -3520 && 0.001 & 0.001 & 0.004 & 0.003 & -1920 \\
$\phi= 41^{\circ}$ &  0 & -13 & 13 & 26 & -3200 && 1.5 & 1.2 & 1.8 & 0.7 & 4800 && 0.005 & 0.007 & 0.015 & 0.008 & 5440 \\
$\phi= 54^{\circ}$ &  0 & -9 &  9 & 18 & -3520 && 2.7 & 2.6 & 3.0 & 0.4 & -4800 && 0.011 & 0.013 & 0.028 & 0.014 & 5760 \\
Fig7(b) $\phi= 24^{\circ}$ &  0 & -25 & 19 & 43 & -7679 && 0.3 & 0.2 & 0.7 & 0.5 & -3520 && 0.001 & 0.001 & 0.003 & 0.002 & -9599 \\
$\phi= 41^{\circ}$ &  0 & -9 & 11 & 19 & -7359 && 1.5 & 1.3 & 2.1 & 0.8 & -4160 && 0.006 & 0.008 & 0.015 & 0.007 & 8639 \\
$\phi= 54^{\circ}$ &  0 & -7 &  8 & 15 & -5440 && 2.7 & 2.8 & 3.2 & 0.5 & -3840 && 0.012 & 0.015 & 0.030 & 0.016 & 8959 \\
Fig7(c) $\phi= 24^{\circ}$ &  0 & -31 & 44 & 75 & -1280 && 0.3 & 0.1 & 0.6 & 0.5 & 1920 && 0.001 & 0.001 & 0.005 & 0.004 & -3200 \\
$\phi= 41^{\circ}$ &  0 & -9 & 25 & 34 & -2880 && 1.5 & 0.9 & 2.1 & 1.1 & 8319 && 0.006 & 0.006 & 0.017 & 0.011 & 8319 \\
$\phi= 54^{\circ}$ &  0 & -5 & 19 & 24 & -2880 && 2.6 & 1.7 & 3.4 & 1.6 & 8639 && 0.012 & 0.012 & 0.028 & 0.016 & 8319 \\
Fig7(d) $\phi= 24^{\circ}$ &  0 & -22 & 29 & 51 &3840 && 0.3 & 0.1 & 0.5 & 0.4 & -4480 && 0.001 & 0.000 & 0.004 & 0.004 & -2880 \\ 
$\phi= 41^{\circ}$ &  0 & -8 & 10 & 18 & -3200 && 1.5 & 1.0 & 2.0 & 1.1 & 4160 && 0.006 & 0.006 & 0.015 & 0.009 & 6719 \\
$\phi= 54^{\circ}$ &  0 & -6 &  6 & 11 & -4480 && 2.7 & 2.3 & 3.6 & 1.3 & 3840 && 0.012 & 0.013 & 0.027 & 0.015 & 7359 \\ 
Fig8(a) $\phi= 24^{\circ}$ &  0 & -12 & 11 & 23 & -1920 && 0.6 & 0.6 & 0.7 & 0.2 & 3520 && 0.002 & 0.002 & 0.006 & 0.004 & 6080 \\
$\phi= 41^{\circ}$ &  0 & -6 &  6 & 11 & -2560 && 1.7 & 1.7 & 1.8 & 0.2 & 3520 && 0.006 & 0.007 & 0.018 & 0.011 & 6719 \\
$\phi= 54^{\circ}$ &  0 & -3 &  3 &  7 & -3520 && 2.4 & 2.5 & 2.7 & 0.2 & 2880 && 0.009 & 0.011 & 0.026 & 0.015 & 6400 \\
Fig8(b) $\phi= 24^{\circ}$ &  0 & -5 &  6 & 11 & -2240 && 0.6 & 0.4 & 0.9 & 0.6 & -3840 && 0.002 & 0.002 & 0.004 & 0.002 & 4480 \\
$\phi= 41^{\circ}$ &  0 & -3 &  3 &  6 & -7039 && 1.7 & 1.5 & 2.1 & 0.6 & 5760 && 0.006 & 0.007 & 0.015 & 0.008 & 7359 \\
$\phi= 54^{\circ}$ &  0 & -2 &  2 &  4 & -6080 && 2.4 & 2.3 & 2.8 & 0.4 & 6719 && 0.010 & 0.012 & 0.026 & 0.014 & 7359 \\
Fig8(c) $\phi= 24^{\circ}$ &  0 & -15 & 19 & 34 & 2880 && 0.6 & 0.4 & 0.9 & 0.5 & 7359 && 0.002 & 0.002 & 0.006 & 0.005 & 7679 \\
$\phi= 41^{\circ}$ &  0 & -8 &  8 & 17 & 3200 && 1.7 & 1.3 & 2.3 & 1.0 & 7999 && 0.006 & 0.007 & 0.017 & 0.010 & 7679 \\
$\phi= 54^{\circ}$ &  0 & -4 &  4 &  9 & 4160 && 2.4 & 2.1 & 3.3 & 1.2 & 8319 && 0.010 & 0.012 & 0.025 & 0.013 & 7359 \\
Fig8(d) $\phi= 24^{\circ}$ &  0 & -12 & 12 & 24 & -2240 && 0.6 & 0.4 & 0.9 & 0.4 & 1920 && 0.002 & 0.002 & 0.005 & 0.004 & 6400 \\
$\phi= 41^{\circ}$ &  0 & -5 &  4 &  9 & -3200 && 1.7 & 1.5 & 2.1 & 0.7 & 4800 && 0.006 & 0.007 & 0.016 & 0.010 & 6400 \\
$\phi= 54^{\circ}$ &  0 & -2 &  2 &  4 & -5760 && 2.4 & 2.3 & 3.1 & 0.8 & 3840 && 0.010 & 0.011 & 0.025 & 0.014 & 7359 \\
&&&&&&&&&&& \\
\hline
\end{tabular}
\end{table*}

The reverse swing can be seen in simple simulations where the
scattering region corresponds to a narrow ring around the BLR and
therefore needs to be explained by simple geometric effects as
follows. Because of the inclined orientation of the scattering region
the near and far regions of the scatterer have a larger amplitude PA
swing (because of projection effects) while PF is low (because these
regions are seen `in transmission' -- see Figure
\ref{schematicfigure}). The opposite is true for the E and W regions
of the scatterer: they are characterized by a smaller amplitude PA
swing and larger PF values. Adding the {\it STOKES\/} parameters of
these two distinct regions of the scatterer will always result in a
reversed PA swing as they combine a larger PA and smaller PF at 90
degrees with a smaller PA and larger PF at 0 degrees\footnote{A PA
  rotation around $\sim 0^{\circ}$ (seen in the E and W) and $\sim
  90^{\circ}$ (seen in the N and S) imply that $Q^{0} \gg U^{0}$ and
  $Q^{90} \gg U^{90}$, but with $Q^{0}$ and $Q^{90}$ having opposite
  signs. The PA spectra at $90^{\circ}$ is also characterized by
  larger amplitude swings (i.e., $U^{90} \gg U^{0}$), which is driven
  by a change from negative to positive values of $U^{90}$ at the line
  center. Since the PF from $0^{\circ}$ dominates, we also have $Q^{0}
  \gg Q^{90}$ and the final $2\times$PA $\sim$ tg$^{-1}
      [(U^{0}\!+\!U^{90})/(Q^{0}\!+\!Q^{90})] \sim$ tg$^{-1}
      [U^{90}/Q^{0}]$. Since $Q^{0}$ has the opposite sign to
      $Q^{90}$, an inverted PA is found.}.

Things are much more complex when an extended geometry and optical
depth effects are also taken into account. Figure \ref{pa_maps} shows
spatially resolved polarization maps for model 2 (d) at two viewing
angle and at wavelengths 6507 and 6612 \AA, which coincide with the
peaks of the PA swings (as marked with vertical lines in Figure
\ref{quad_PA_spectra}). It can be seen that the projection effects
that determine the amplitude of the PA swing become more severe the
larger the viewing angle. Figure \ref{ring_spec} shows the spectra of
the polarized flux (PF) obtained from concentric and projected rings
defined on the plane of the sky, which are also shown in Figure
\ref{pa_maps}. Note that due to the very thick scatterer (height
$\sim$ radius -- see Table 1), polarized flux at $\sim 90$ degrees is
observed coming from the high walls of the scattering region in the
near and far sides, while absent in the orthogonal regions.

The PF spectra show no clear progression in the strength of the
polarized flux from ring to ring, which is counter-intuitive. This is
the result of a combination of optical depth and geometric
cancellation. In fact, further tests with much lower optical depths
demonstrate that a clear pattern is recovered for PF as a function of
radius: regions at intermediate radii show the strongest polarized
flux with the strength diminishing towards the center and the edge of
the scattering disk.

Still, despite all this complexity, in the case of a simple scattering
ring the reasoning at the beginnning of this section can explain why
the PA swing for model 2 (d) is reversed compared with that proposed
by S05.  The polarized signal coming from the N and S regions of the
scatterer cannot be simply dismissed, but needs to be combined with
the {\it STOKES\/} fluxes from the dominant E and W regions in order
to find the final solution.

\begin{figure*}
\includegraphics[scale=0.85,trim=0 0 0 0]{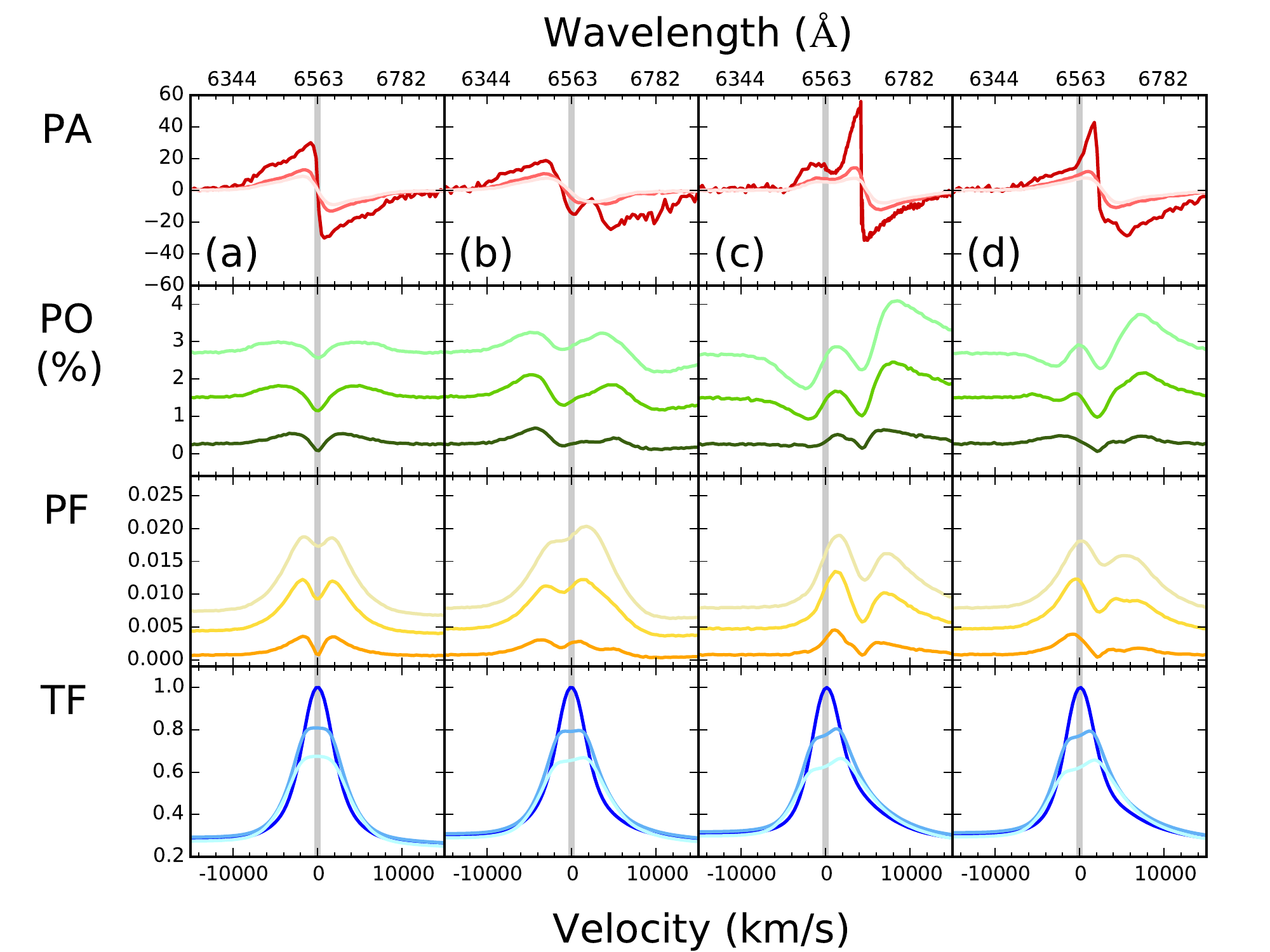}
\caption{{\it STOKES\/} modelling for a Keplerian rotating equatorial
  outer scatterer (a), undergoing a vertical outflow (b), a radial
  outflow (c), and a 45 degree inclined outflow (d). All models
  consider emission from the central source. From top to bottom, we
  show the position angle (PA), percentage polarization (PO),
  polarized flux (PF) and total flux (TF). From darker to lighter
  shades, the three models shown correspond to viewing angles of 24,
  41 and 54 degrees measured from the axis of symmetry of the system.}
\label{modelKep}
\end{figure*}

\begin{figure*}
\includegraphics[scale=0.85,trim=0 0 0 0]{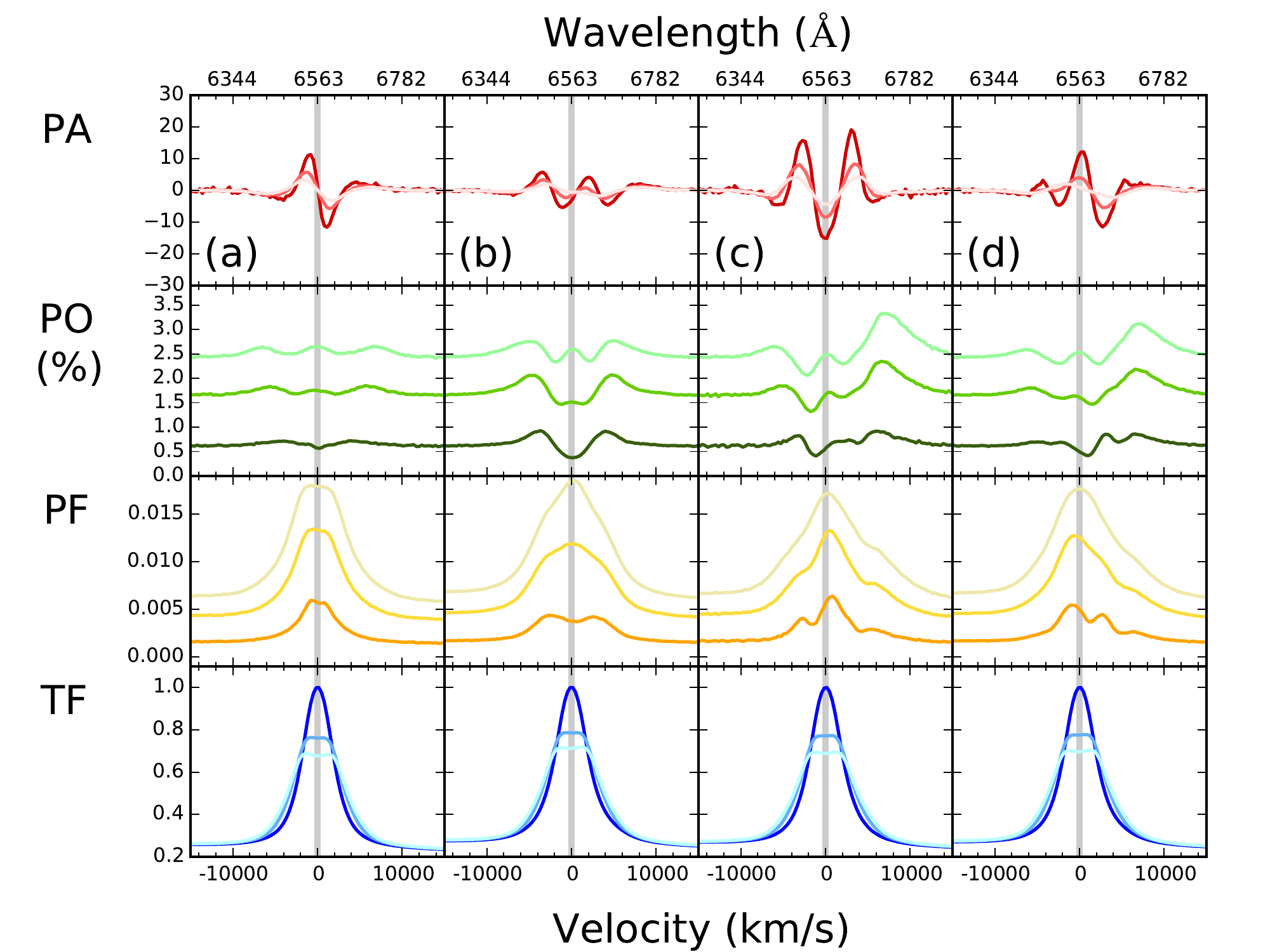}%
\caption{{\it STOKES\/} modelling for a Keplerian rotating scatterer
  coincident with the location of the disc-like BLR (a), undergoing a
  vertical outflow (b), a radial outflow (c), and a 45 degree inclined
  outflow (d). All models consider emission from the central
  source. From top to bottom, we show the position angle (PA),
  percentage polarization (PO), polarized flux (PF) and total flux
  (TF). From darker to lighter shades, the three models shown
  correspond to viewing angles of 24, 41 and 54 degrees measured from
  the axis of symmetry of the system.}
\label{modelBLR}
\end{figure*}

\section{Scatterers in Keplerian rotation}

Figure \ref{modelKep}, Panel (a), presents a model including Keplerian
rotation of the scattering medium for the same parameters presented in
Figure \ref{modelS05}, Panel (e). As the scatterer is located at
larger radii than the BLR, and $v_{\phi}$ goes as the inverse of the
square root of the radius, the resulting relative velocity between the
two structures is in the same direction as before but with smaller
magnitudes. Hence, the scattered line photons present a smaller
velocity gradient than in the static configuration. On the other hand,
as the scattering region is also rotating with respect to the
observer, the photons acquire a further velocity shift after the
scattering event. The magnitude of this shift depends on the azimuthal
position of the scattering element with respect to the observer. For
elements at the near and far side of the scattering region, no shift
is introduced as the scatterer is at rest as seen by the observer.
Note that these scattering elements, however, contribute little to the
scattering signal. For elements at the orthogonal positions, which
dominate the observed scattering, the shift is equal to the Keplerian
velocity, which corresponds to up to $\sim 2000$ km/s (see Table 1 and
discussion below).

Some differences appear at the center of the line when comparing the
spectra in Figures \ref{modelS05} Panel (e) and \ref{modelKep} Panel
(a), because of the smaller velocity differences between the BLR and
scatterer. In particular, the switch in PA values occurs at a smaller
velocity range and with larger amplitude, resulting in a sharp swing
(see Table 2). At the same time, the wings of the PA profile look more
extended because of the velocity shift experienced by the photons at
the orthogonal positions: while the W side of the scatterer recedes
from us, the E side is approaching us. This also produces small humps
in the wings, which are caused by the scattering of the innermost BLR
photons: a sharp PA swing at large velocities is produced because of
the small subtended angles. This PA shape is smoothed at the
orthogonal positions because of the extra velocity shifts, while
staying sharp in the signal coming from the near and far sides of the
scatterer. The presence (or absence) of these features is therefore
dependent on the velocity ranges shown by the BLR and scatterer.

Results are also shown in Figure \ref{modelKep} when including 3000
km/s outflowing ionized material in the vertical ($v_{z} = \pm 3000$
km/s -- Panel (b)), radial ($v_{\rho} = 3000$ km/s -- Panel (c)), and
at a 45 degrees ($|v_{z}| = v_{\rho} = 3000/\sqrt 2$ km/s -- Panel
(d)). Geometric and dynamical constraints can be found in Table
1. When the scatterer presents a net velocity with respect to the BLR
two Doppler shifts are involved in any scattering event. The first
shift occurs when switching from the reference frame of the BLR (or
the previous scattering electron) into the frame of the scattering
material. The second occurs after the scattering event, when
transforming to the frame of the observer (or the following scattering
electron).

The choice of 3000 km/s for the outflowing material is motivated by
the velocities reached by the rotating scatterer. For the parameters
adopted for this region (Keplerian rotation around a $\sim 10^7$
M$_{\sun}$ black hole with the innermost orbit located at $\sim 0.04$
pc), the equatorial scatterer reaches velocities of $\sim 2000$
km/s. In order for the outflowing component to introduce significant
changes its velocity must be of a similar or larger magnitude.

The vertical outflows shown in Figure \ref{modelKep} Panel (b) are
launched into both hemispheres, i.e., $v_{z} = \pm 3000$ km/s. To
first approximation there is very little net velocity shift between
the BLR and the scatterer since the outflowing motion is mostly
orthogonal to the BLR motion. This translates into profiles without
any net shifts. However, the PA profile becomes complex because the
outflow has a component that moves towards us and one that recedes
from us, introducing a double swing in the scattering signal and some
suppresion in the PA amplitude.  Obviously the effects of the outflow
are stronger for small inclinations angles, which maximize the
velocity outflowing component along the line of sight towards the
observer. At larger inclinations the PA profile looks much closer to
that seen in Figure \ref{modelKep} Panel (a).

The PF profile in Panel (b) appears broader than in Panel (a) due to
the velocity shifts introduced by the outflowing scatterer. The
profile presents an enhanced red horn at large viewing angles (54
degrees) and an enhanced blue horn at small viewing angles (24
degrees), but it does not present a net velocity shift as the
dynamical configurations from the point of view of the scattering
medium and the observer are symmetric which, as we will see, it is not
the case with the next model.

Panel (c) in Figure \ref{modelKep} depicts the results from having
radial motions. The profiles are clearly shifted redwards as the
scatterer is moving away from the BLR in the meridional plane. Also,
the effects are stronger than in the case of the vertical outflow
because the velocity shifts preferentially occur in the same plane as
the travel direction of the scattered photons. The amplitude and
sharpness of the PA swing in the less inclined line of sight are both
dramatically increased. In the PF spectrum a broad double horn profile
of asymmetric intensity appears. Also, the N-S asymmetry discussed in
Section 3.2 implies that the redshifted emission from the far side
dominates over the near side blushifted emission. This produces a red
tail visible beyond 10000 km/s.

Finally, Panel (d) presents the case of a wind outflowing at 45
degrees with respect to the meridional plane of the system. The
results are a combination of those already seen in Panels (b) and
(c). The PA amplitude variation is suppressed due to the vertical
component of the scatterer velocity, while the radial component
introduces a redshift to the line features which is smaller than that
seen in Panel (c).

S05 also considered the effects of a radial 900 km/s inflowing,
non-rotating wind and their results can be compared with our Panel (c)
in Figure \ref{modelKep}. Other than the obvious net blue-shift in the
lines, as the scattering material is inflowing towards the BLR in the
S05 models, the main difference with their results is the symmetric
peaks in the PF spectra, because S05 considered single scattering
events. Also, S05 did not find the strong asymmetric PA profile we see
for low inclination angles. This might be due to their outflowing
velocity, which is not fast enough to significantly change the
velocity structure of the outflowing material.

\section{Coincident scattering and emitting regions}

A variation of the S05 model has the rotating scattering medium near
the BLR location. This could be acieved in two ways.

First, for a thin layer located above and below, sandwiching the BLR
region, photons escaping vertically from the emitting material will
undergo no scattering, while photons escaping at grazing angles
encounter a much larger optical depth and will be scattered. Hence the
net PA will be similar to that of the system axis. Second, for a thin
scatterer spatially coincident with the BLR region, the same mechanism
applies, and a parallel PA is also recovered. These two model
realizations (the sandwich and the coincident scatterer) give
virtually identical polarization signatures. We present the results
corresponding to the coincident scatterer in Figure \ref{modelBLR}.

Note that the thickness of the BLR has very little impact on our
results, but the thinness of the scattering medium is crucial to
produce the desired parallel polarization signal. Following S05, a
thin BLR has been used in all our models, although it is not clear
whether this is a good representation of this structure (see
references in the Introduction). For the modelling of a BLR coincident
with the scattering region, however, a thin BLR / scatterer geometry
must be used since {\it STOKES\/} cannot represent the scatterer as a
layer surrounding individual clouds.

Two important differences should be noted in the case of a coincident
BLR and scattering region. First, because the scatterer will scatter
photons `locally' (i.e., not in rings of scattering material located
outside the BLR), there will be no N-S asymmetry due to the large
optical depth of the scattering region (see Section 3.1). Second,
because the scattering elements will be bombarded by photons coming
from all directions, photons coming from directions perpendicular to
our line of sight will yield the highest polarization level (see
Figure \ref{schematicfigure}), while those approaching the scattering
material in directions parallel to our line of sight will give a
rather negligible, `in transmission', polarization signal. Before,
geometric cancellation happened when combining the signal from
opposite regions of the scatterer; now, cancellation will become
important at all positions.

\begin{figure*}
\includegraphics[scale=0.42,trim=100 200 0 200]{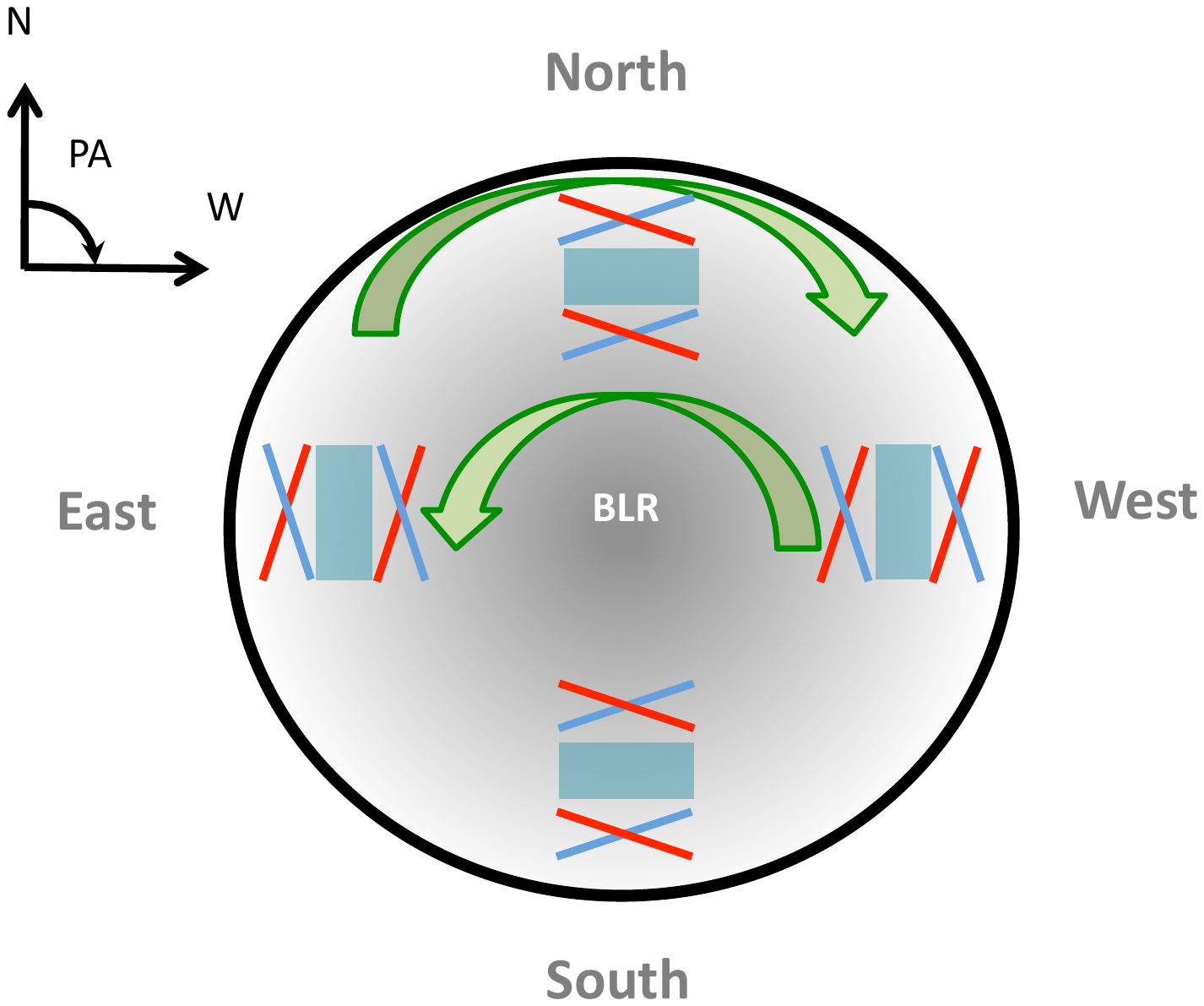}%
\includegraphics[scale=0.25,trim=150 0 0 0]{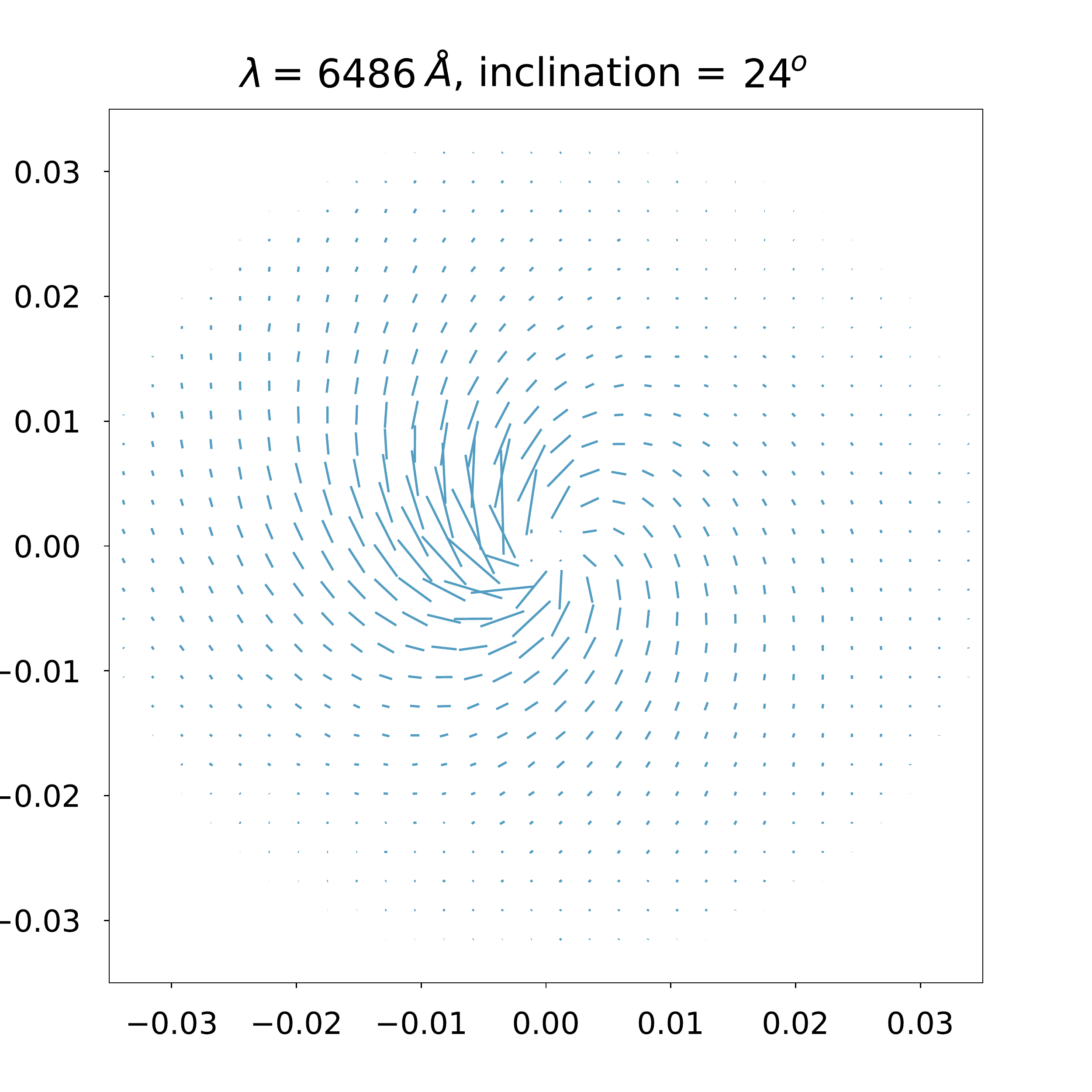}%
\includegraphics[scale=0.25,trim=50 0 0 0]{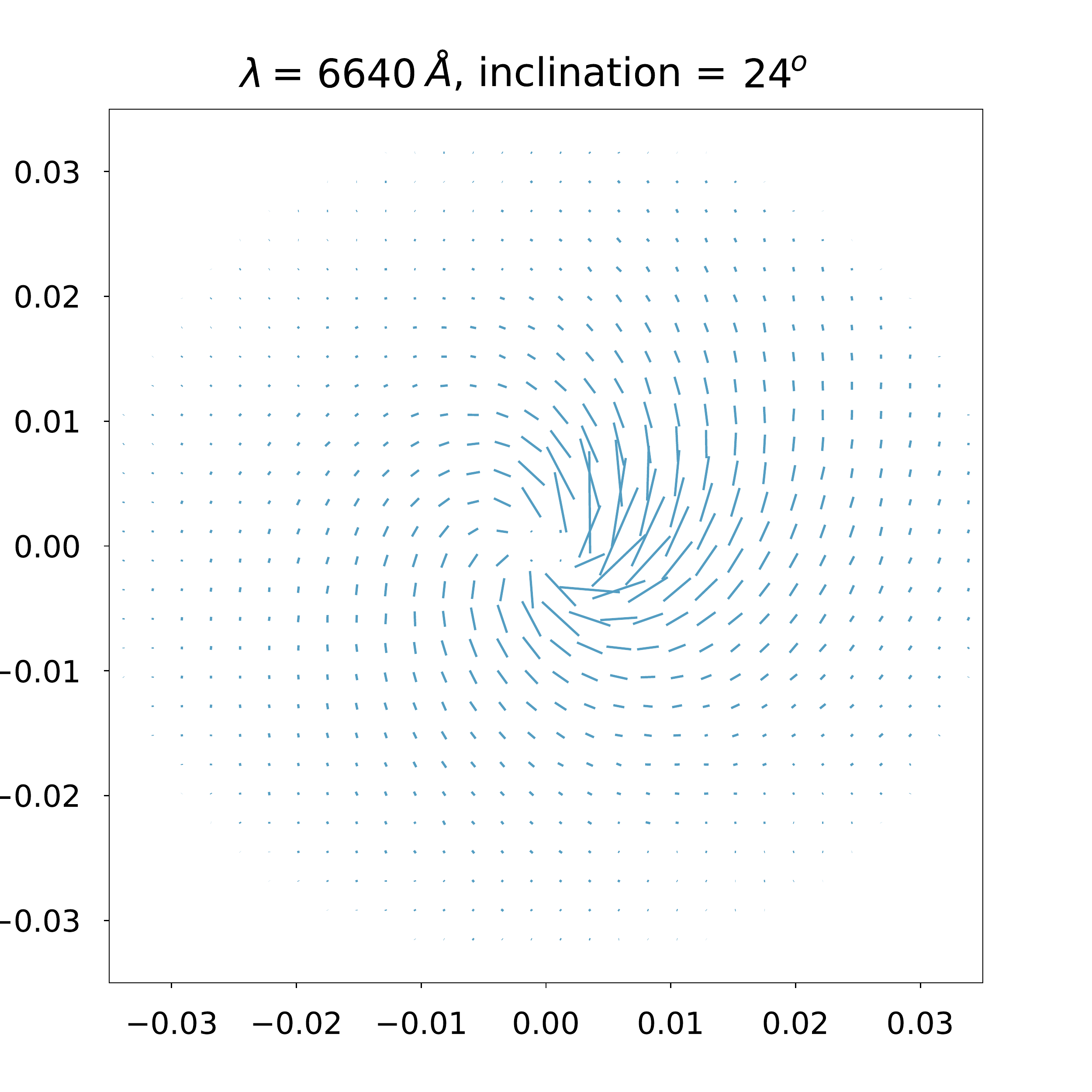}
\caption{{\bf Left:} Schematic representation of the position angles
  seen by elements of the scattering medium for a scatterer coincident
  with the emitting BLR and in Keplerian rotation. {\bf Center and
    right:} PA spatially resolved maps of model 8(a) for wavelengths
  6486\AA\ and 6640\AA. The sizes of the markers represent the
  strengths of the polarized fluxes PF and their inclinations
  represent the PA angles. The axes are expressed in pc.}
\label{pa_maps_2}
\end{figure*}

The scattering events will typically occur when the optical depth of
photons reaches a value $\sim 1$. For the electron density assumed in
our models this will happen after the photons have travelled about
0.015 pc in a trajectory nearly parallel to the disc mid-plane. This
corresponds to $\sim 1/3$ of the disc diameter, which defines the term
`locally' used in the previous paragraph. For those photons produced
in the innermost regions of the disc scattering will take place in a
wide, nearly annular region found at the radius where $\tau \sim 1$,
not too differently from the geometry presented in the previous
sections. For photons produced in the middle regions of the disc
things will differ significantly as they can travel inwards and
outwards until they encounter an electron and undergo a scattering
event, as discussed further below. Photons produced in the outermost
regions of the disc will contribute rather little to the scattering
flux as a large fraction of their emission will escape the system. A
first approximation, therefore, is to consider only the regions of the
BLR found at small and medium radii. In the frame of a scaterrer, due
to the Keplerian rotation of the disc, regions at smaller radii will
rotate in an anti-clockwise manner, while those at larger radii will
rotate in the opposite direction. So approaching and receding photons
will be scattered at the same PA irrespective of whether the photon
originated at smaller or larger radii. This is schematically shown in
the left panel of Figure \ref{pa_maps_2}.

At intermediate radii the polarized signal is not produced by photons
approaching the scatterer radially, but by those that have travelled a
distance such that $\tau \sim 1$, which is satisfied at positions
offset from the center. Hence, at the blue side of the line the N
region will essentially produce scattering of those photons coming
from the W, while the W region will produce scattering of photons
coming from the S, and so on, and a swirling pattern will emerge for
the polarization signal. At the red side of the line, the pattern will
rotate in the opposite direction, as can be seen in Figure
\ref{pa_maps_2}. These patterns will dissappear as the disc becomes
more and more inclined, since photons coming from directions
perpendicular to our line of sight will dominate the signal.

\begin{figure*}
\includegraphics[scale=0.35,trim=0 0 0 0]{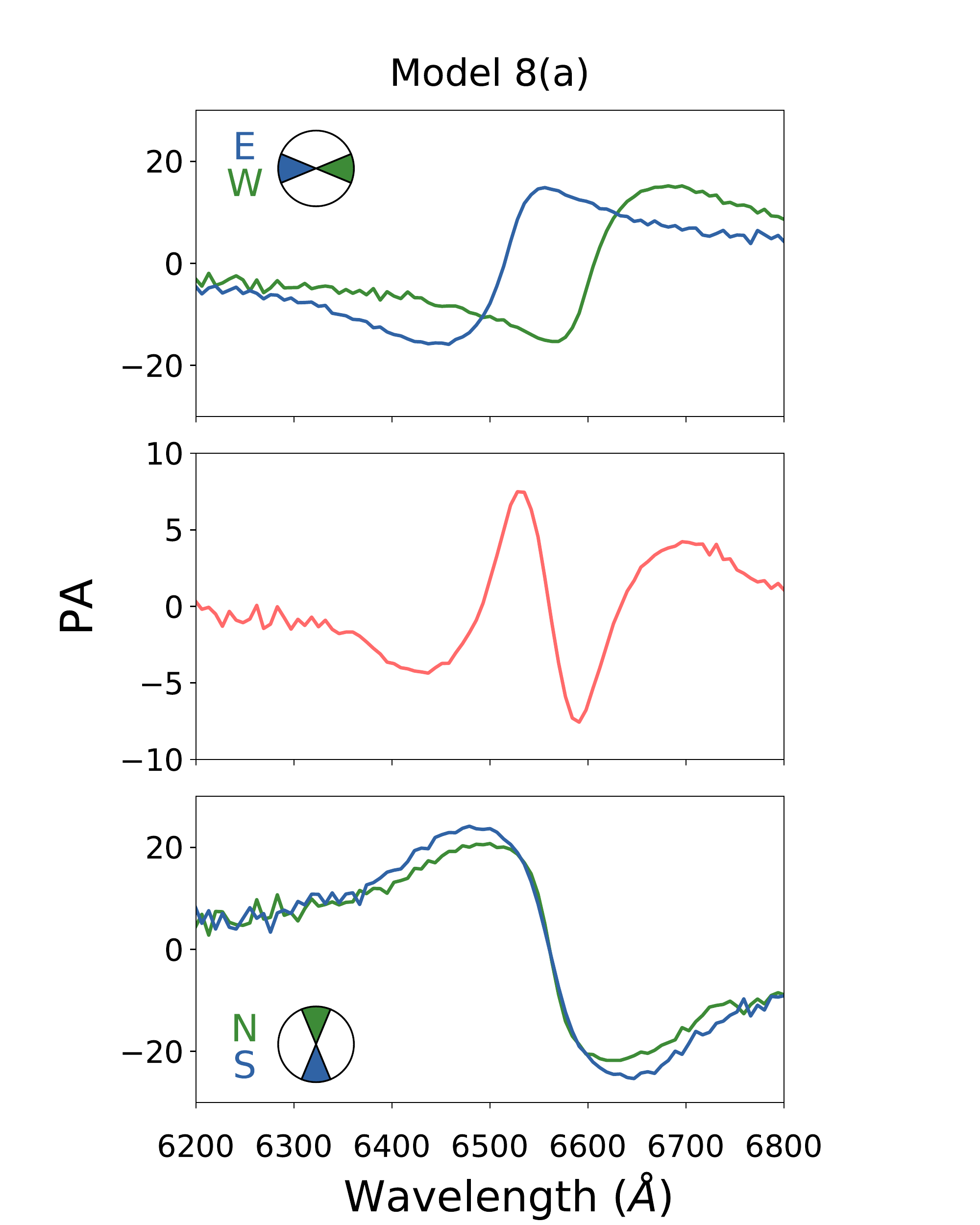}%
\includegraphics[scale=0.35,trim=100 0 0 0]{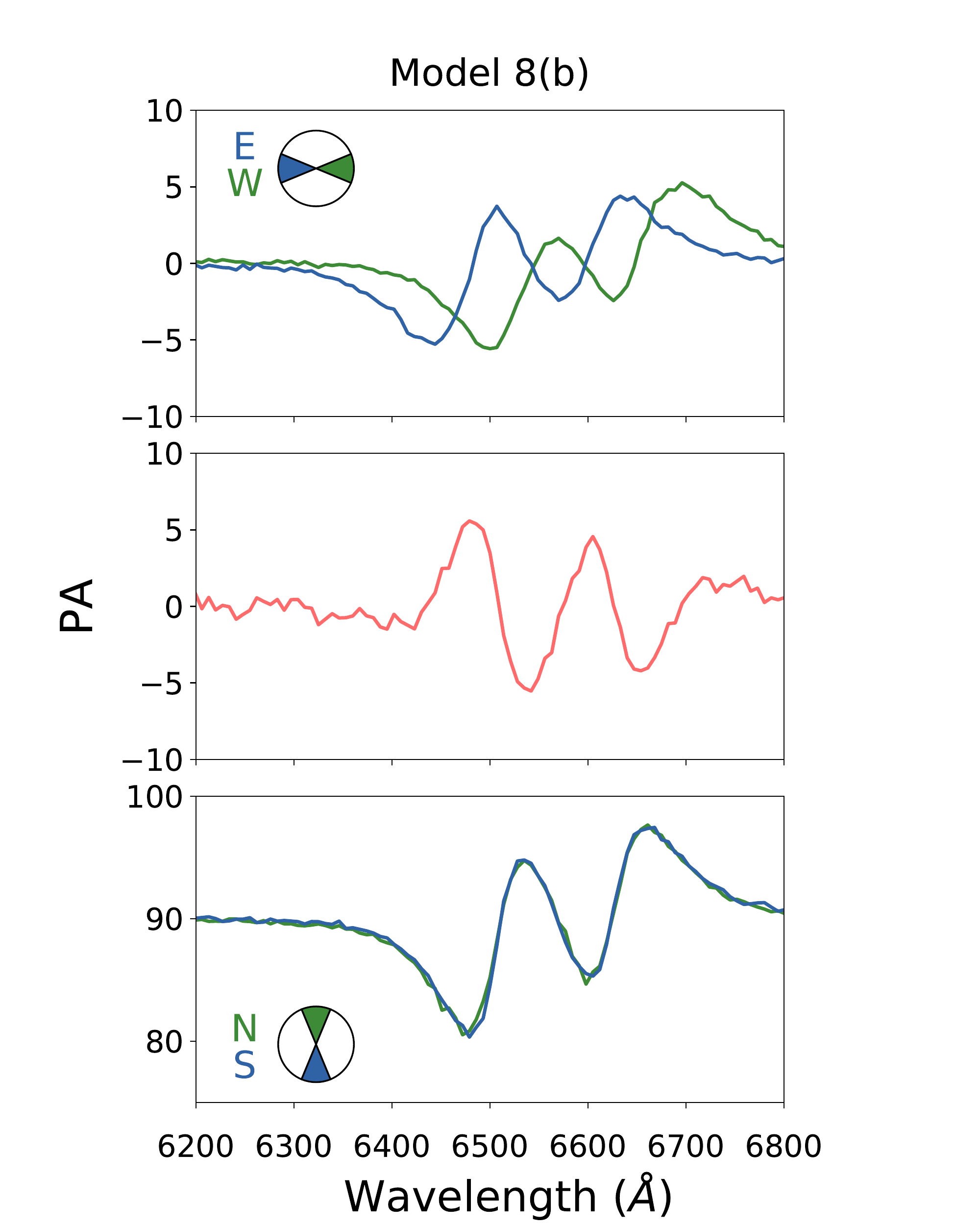}%
\includegraphics[scale=0.35,trim=100 0 0 0]{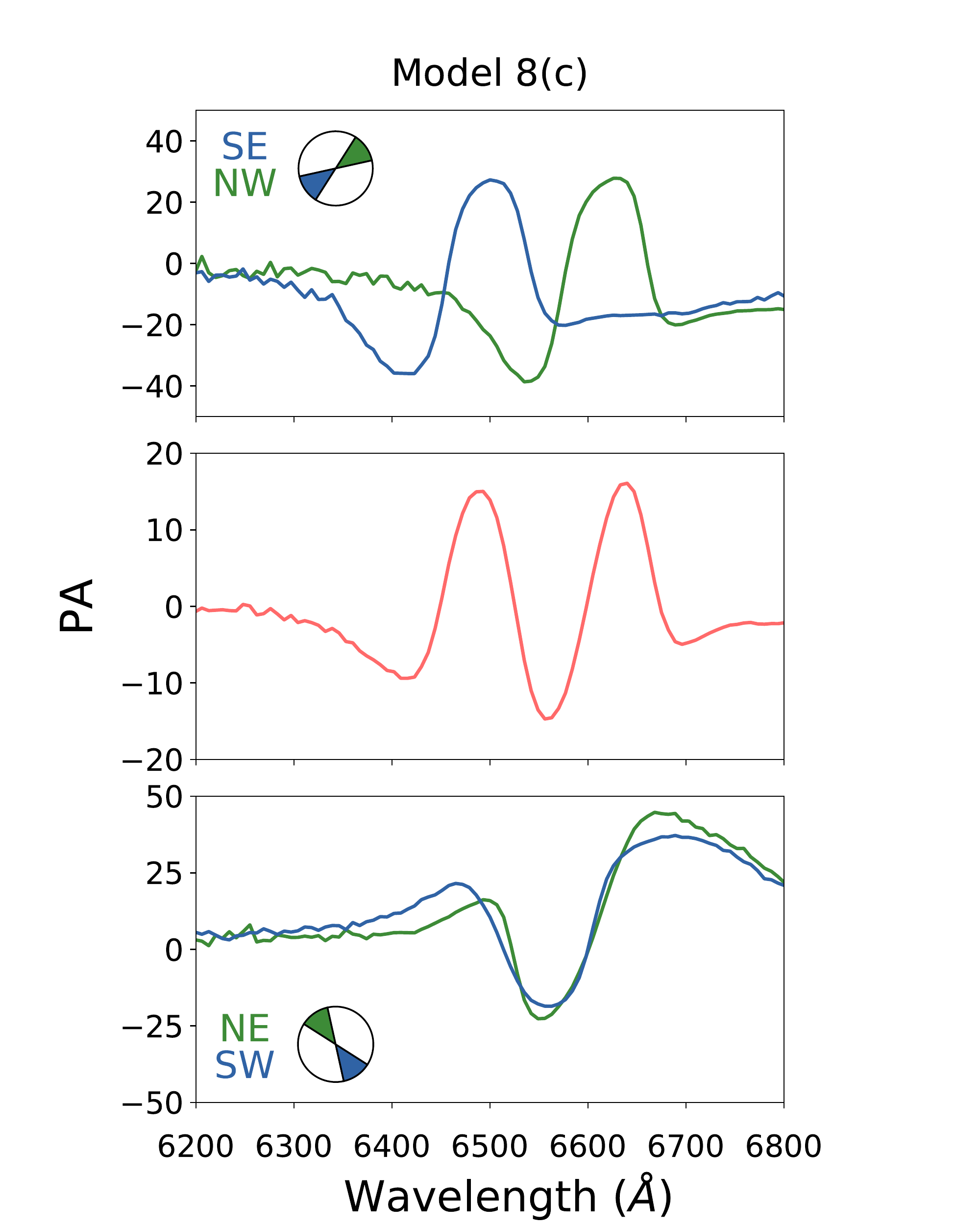}
\caption{Detailed PA spectra for models presented in Figure 8(a), 8(b)
  and 8(c). Top and bottom panels show the PA spectra obtained from
  $\pi/8$ wedge-like regions, as depicted in the pie-charts. The
  central panels shows in red the total PA spectrum obtained for the
  full scattering region. For further details see text.}
\label{spec_pie}
\end{figure*}

In the absence of outflowing motions (Figure \ref{modelBLR}, Panel
(a)) to the previously described effects we only need to add velocity
shifts from the rotation of the scatterer as seen by the observer,
with the signal from the E side of the disc being blue-shifted while
that from the W side of disc is redshifted. The N and S quadrants, on
the other hand, present negligible shits, as can be seen in the left
panel of Figure \ref{spec_pie}. Going from shorter to longer
wavelengths (from the 6486\AA\ to the 6640\AA\ PA maps presented in
Figure \ref{pa_maps_2}), the PA signal in the E and W quadrants
rotates clockwise, while that in the N and S rotates anticlockwise, as
expected. Hence, while the E and W quadrants present a PA profile like
the one sketched in Figure \ref{schematicfigure}, the S and N do the
opposite, as can be seen in Figure \ref{spec_pie}. The combined signal
corresponds to the central swing seen in the final PA profile, with
the regions with the largest polarized flux seen in Figure
\ref{pa_maps_2} dominating.

Note the low amplitude variations seen across the line in the PO
spectrum in Figure \ref{modelBLR} Panel (a), while the PF spectrum is
rather narrow and shows very little evidence for a double horn
profile. These are the result of the scattering medium having the same
Keplerian velocity as the BLR and the scattering elements seeing the
photons from approaching and receding BLR regions at very low relative
azimuthal velocities.

Models with the same set of outflow velocities previously discussed
(vertical, radial and 45 degree inclined) are also shown in
\ref{modelBLR}. Note that the scattering events occur close to the BLR
disk despite the different outflowing velocities of the scattering
material. In other words, despite the fact that the electrons might be
flying away from the BLR, it is only while they belong to a dense thin
layer of material that they are able to produce scattering. By not
modelling the presence of this wind far away from the BLR, we assume
that no further interaction of the photons with these outflowing
electrons takes place.

The presence of vertical outflowing motion in the scattering region
introduces changes to the polarized spectra as seen in Panel (b). The
outflow introduces secondary peaks as the original sinusoidal profiles
are blue and red-shifted by the wind (as seen in the center panel of
Figure \ref{spec_pie}), which for small and intermediate disc
inclinations -- expected for Seyfert I nuclei -- will have a
significant component along the line of sight to the observer.
Rotation will affect the E and W regions to introduce yet another
shift, also seen in Figure \ref{spec_pie}. The PF spectra in Figure
\ref{modelBLR} look much broader in the presence of a wind, as the
approaching and receding velocity components of the outflowing
scatterer introduce large Doppler shifts.

Again things appear significantly different when the scattering medium
has radial outflows. Figure \ref{modelBLR} Panel (c) shows a
symmetric, `M-shaped' PA and large amplitude swing which is in stark
contrast to Figure \ref{modelKep}. The explanation for these new
profiles is analogous to that presented earlier for Panels (a) and
(b).

In the rest frame of the scattering elements, there will be large
velocity offsets with respect to the emitting BLR due to the radial
outflow. At each location the PA profile will combine the blue-shifted
swing from the BLR located at larger radii with a red-shifted swing
from the BLR located at smaller radii. A combined PA profile
corresponding to a complete back-and-forth rotation will emerge.

To further understand this case it is best to split the scatterer into
NE-NW-SE-SW regions. As seen by the observer, the combination of the
radial outflow and the rotation motion will give NW and SE regions
receding and approaching, respectively, while the NE and SW regions
will remain largely at rest. Hence the SE and NW PA profiles will get
pushed blue and red-wards, respectively, as can be seen in right panel
of Figure \ref{spec_pie}, while the NE and SW profiles stay at the
rest line velocity. Back-and-forth PA rotations will occur in a
clockwise followed by an anti-clockwise manner for the NW and SE
regions, while an anti-clockwise rotation followed by a clockwise
rotation will occur at the NE and SW regions, as seen in Figure
\ref{spec_pie}. The final combination of all regions gives a very
symmetric M-shaped PA profile.

Note that, as before, projection effects will affect all velocity
components in the same way, and therefore the emerging profile is
symmetric in velocity space. The only differences would appear because
of possible boosting effects along the line of sight. However, as all
velocity fields are contained in the plane of the disc, and in Seyfert
I nuclei this inclination angle of the disc is usually small, boosting
should not be significant for this case.

Panel (d) presents the case for a 45 degree outflow. In this case, the
features already seen in cases (b) and (c) play a role. The final PA
profile is complex and asymmetric.

\section{Discussion}

\subsection{General results}

Using state of the art Monte Carlo {\it STOKES\/} simulations we have
been able to recover and extend the work done by S05 on the polarized
signal of broad emission lines in Type I AGN. Our treatment finds
similar results to those published by S05, but requires some
significant changes to the S05 original set up, as discussed in
Section 3. We found that the final scattering signal cannot be based
solely on emission from the orthogonal sides of the scatterer;
emission from the near and far sides must be included to find the
correct polarized signal. As a result, our treatment yields a reverse
PA swing to that originally proposed. Also, a much larger optical
depth (up by one order of magnitude) is required to obtain significant
levels of polarization (a few \%), and a much thicker equatorial
scatterer is needed to recover the emission line PA swing discussed at
length by S05.

Marin et al.~(2012, 2013) also found that large optical depths ($\tau
\sim 1-3$) are necessary to obtain the level of polarization usually
seen in AGN. Using {\it STOKES\/} to model the phenomenologically
motivated central source structure proposed by Elvis (2000), Marin et
al.~showed that an outflowing electron scatterer will yield
polarization of a few percent if the wind is optically thick to
electron scattering.

We have included the presence of the central continuum emitting source
as part of our modelling, which clearly represents a more accurate
prescription of the scattering signal. The inclusion of large (3000
km/s) bulk motion due to an outflowing scatterer shows that the
scattering signal can deviate significantly from the S05 predictions,
yielding new profiles that can be readily tested against observations.

We also explored a new configuration for the BLR and scatterer regions
where they are spatially coincident, either with the scatterer acting
as an `atmosphere' for the BLR or fully mixed with it. The results
from this new configuration give very different results when outflows
are considered, with an M-shaped PA profile being predicted as a
result of the scattering of photons undergoing symmetric shifts from
the wind and rotational fields. This M-like signal is a clear
indication of the scattering taking place in a cospatial BLR and
equatorial outflow and has already been observed in
spectropolarimetric observations of nearby Seyfert galaxies, as
discussed below.

\subsection{Qualitative comparison with observations}

High signal-to-noise ratio spectropolarimetric observations of Seyfert
I galaxies have become available in recent years (e.g., Afanasiev et
al.~2019). Some of the observations show rich PA profiles that could
not be explained by the S05 results: NGC~3783 spectropolarimetric
observations do not present the usual sinusoidal swing, but a deep
M-like shaped morphology instead (Lira et al.~2007). We believe the
observed profile can be explained by the cospatial nature of the BLR
and an outflowing equatorial scatterer.  Detailed simulation of these
sources is left to a future paper.

\section*{Acknowledgments}

PL and RG acknowledge financial support from the French-Chilean CNRS
UMI. PL acknowledges funding from Fondecyt Project \#1161184 and
partial support from Center of Excellence in Astrophysics and
Associated Technologies (PFB 06). MK acknowledges funding from JSPS
under grant 16H05731. Finally, PL acknowledges the help of
Dr.~A.~Cooke for proof-reading the manuscript.

\appendix

\bsp

\label{lastpage}

\end{document}